\documentclass[11pt]{article}
\pdfoutput=1

\usepackage{cite}
\usepackage{booktabs}
\usepackage[english]{babel}
\usepackage{mathtools}
\usepackage{amsmath,amssymb,amsbsy,amstext, amsthm, simplewick, amsfonts}
\usepackage{hyperref}
\usepackage{graphicx}
\usepackage{amsfonts}
\usepackage{amssymb}
\usepackage[small]{caption}
\usepackage{upgreek}
\usepackage[svgnames,dvipsnames,x11names,table]{xcolor}
\usepackage{multirow}
\usepackage{geometry}
\usepackage[hang,flushmargin]{footmisc}
\usepackage{bm}
\usepackage{braket}
\usepackage{subcaption}
\usepackage{mathtools}
\usepackage{setspace}
\usepackage{cleveref}
\usepackage{comment}
\usepackage{scalerel}
\usepackage[normalem]{ulem}
\usepackage{slashed}
\usepackage{tensor}
\usepackage{enumitem}
\usepackage{ifthen}
\usepackage{appendix}
\usepackage{floatrow}
\usepackage{siunitx}
\usepackage{framed}
\usepackage{float}
\usepackage{geometry}
\usepackage{fontawesome}
\usepackage{tikz}
\usetikzlibrary{decorations.markings}
\usetikzlibrary{shapes.misc}
\usetikzlibrary{arrows}
\usetikzlibrary{shapes.geometric}
\usetikzlibrary{plotmarks}

\makeatletter
\g@addto@macro\bfseries{\boldmath}
\makeatother

\numberwithin{equation}{section} 

\renewcommand{\thefootnote}{\fnsymbol{footnote}}

\newcolumntype{M}[1]{>{\centering\arraybackslash}m{#1}}
\definecolor{pyblue}{RGB}{31, 119, 180}
\definecolor{pyorange}{RGB}{255, 127, 14}
\definecolor{pygreen}{RGB}{44, 160, 44}
\definecolor{pyred}{RGB}{214, 39, 40}
\definecolor{lightgray}{gray}{0.9}

\hypersetup{
    colorlinks=true,
    linkcolor={Purple!80!black},
    citecolor={pyblue},
    urlcolor={Purple!50!black}
}

\usepackage{colortbl}

\makeatletter
\newlength{\apb@width}
\newcommand{\autoparbox}[2][c]{\settowidth{\apb@width}{#2}\parbox[#1]{\apb@width}{#2}}

\makeatother

\usepackage[framemethod=default]{mdframed}
\newmdenv[skipabove=7pt,
skipbelow=7pt,
rightline=false,
leftline=false,
topline=false,
bottomline=false,
backgroundcolor=gray!10,
linecolor=gray,
innerleftmargin=5pt,
innerrightmargin=5pt,
innertopmargin=5pt,
innerbottommargin=5pt,
leftmargin=0cm,
rightmargin=0cm,
linewidth=4pt]{eBox}

\usepackage[most]{tcolorbox}
\tcbset{colback=white, colframe=black,
        highlight math style= {enhanced, 
            colframe=red,colback=red!10!white,boxsep=0pt}
        }

\crefname{table}{Table}{Tables}
\crefname{equation}{Eq.}{Eqs.}
\crefname{appendix}{App.}{Apps.}
\crefname{section}{Section}{Secs.}
\crefname{figure}{Fig.}{Figs.}

\def \d {\mathrm{d}}
\def \K {\mathcal{K}}
\def \G {\mathcal{G}}
\def \A {\mathcal{A}}
\def \S {\mathcal{S}}
\def \L {\mathcal{L}}

\def \D {\mathcal{D}}
\def \C {\mathcal{C}}
\def \N {\mathcal{N}}
\def \I {\mathcal{I}}
\def \Li {\text{Li}}
\def \O {\mathcal{O}}
\def \Disc {\text{Disc}}

\newmuskip\pFqmuskip
\newcommand*\pFq[6][8]{%
  \begingroup 
  \pFqmuskip=#1mu\relax
  \mathcode`\,=\string"8000
  \begingroup\lccode`\~=`\,
  \lowercase{\endgroup\let~}\pFqcomma
  {}_{#2}F_{#3}{\left[\genfrac..{0pt}{}{#4}{#5};#6\right]}%
  \endgroup
}
\newcommand{\pFqcomma}{\mskip\pFqmuskip}

\newcommand*\ptFq[6][8]{%
  \begingroup 
  \pFqmuskip=#1mu\relax
  \mathcode`\,=\string"8000
  \begingroup\lccode`\~=`\,
  \lowercase{\endgroup\let~}\pFqcomma
  {}_{#2}\tilde{F}_{#3}{\left[\genfrac..{0pt}{}{#4}{#5};#6\right]}%
  \endgroup
}

\newcommand*\pcalFq[6][8]{%
  \begingroup 
  \pFqmuskip=#1mu\relax
  \mathcode`\,=\string"8000
  \begingroup\lccode`\~=`\,
  \lowercase{\endgroup\let~}\pFqcomma
  {}_{#2}\mathcal{F}_{#3}{\left[\genfrac..{0pt}{}{#4}{#5};#6\right]}%
  \endgroup
}

\allowdisplaybreaks[1]
\begin{document}

\begin{titlepage}
\setcounter{page}{1} \baselineskip=15.5pt
\thispagestyle{empty}
$\quad$
\vskip 0 pt

\vspace*{0cm}

\begin{center}
{\fontsize{18}{18} \bf Spectral Representation of}\\[10pt] 
{\fontsize{18}{18} \bf  Cosmological Correlators}
\end{center}

\vskip 15pt
\begin{center}
\noindent
{\fontsize{12}{18}\selectfont Denis Werth\footnote[2]{\href{werth@iap.fr}{werth@iap.fr}}}
\end{center}

\begin{center}
\textit{Institut d'Astrophysique de Paris, Sorbonne Université, CNRS, Paris, F-75014, France}
\end{center}

\vspace{0.4cm}
\begin{center}{\bf Abstract}
\end{center}

Cosmological correlation functions are significantly more complex than their flat-space analogues, such as tree-level scattering amplitudes. While these amplitudes have simple analytic structure and clear factorisation properties, cosmological correlators often feature branch cuts and lack neat expressions. In this paper, we develop off-shell perturbative methods to study and compute cosmological correlators. We show that such approach not only makes the origin of the correlator singularity structure and factorisation manifest, but also renders practical analytical computations more tractable. Using a spectral representation of massive cosmological propagators that encodes particle production through a suitable $i\epsilon$ prescription, we remove the need to ever perform nested time integrals as they only appear in a factorised form. This approach explicitly shows that complex correlators are constructed by gluing lower-point off-shell correlators, while performing the spectral integral sets the exchanged particles on shell. Notably, in the complex mass plane instead of energy, computing spectral integrals amounts to collecting towers of poles as the simple building blocks are meromorphic functions. We demonstrate this by deriving a new, simple, and partially resummed representation for the four-point function of conformally coupled scalars mediated by tree-level massive scalar exchange in de Sitter. Additionally, we establish cosmological largest-time equations that relate different channels on in-in branches via analytic continuation, analogous to crossing symmetry in flat space. These universal relations provide simple consistency checks and suggest that dispersive methods hold promise for developing cosmological recursion relations, further connecting techniques from modern scattering amplitudes to cosmology.

\end{titlepage}

\renewcommand*{\thefootnote}{\arabic{footnote}}
\setcounter{footnote}{0}

\setcounter{page}{2}

\restoregeometry

\begin{spacing}{1.2}
\newpage
\setcounter{tocdepth}{3}
\tableofcontents
\end{spacing}

\setstretch{1.1}
\newpage

\section{Introduction}

The complexity of real physical processes is often made simple when these processes are extended to the complex plane. While thinking about particles propagating with complex momenta and scattering at imaginary angles may seem grotesque, this perspective can offer profound insights into the real world. 
Even more remarkably, fundamental physical principles themselves can be read off from analytic properties of scattering amplitudes. 

\vskip 4pt
The development of off-shell methods for scattering amplitudes is deeply rooted in the power of complex analysis. This approach traces its origins back to the celebrated Kramers-Kronig relations, which link absorption and dissipation in a medium by relating the real and imaginary parts of the response function, thereby establishing the foundational connection between causality and analyticity. Building on this foundation, the complex deformation of momenta has led to significant developments, such as Berends-Giele recursion relations~\cite{Berends:1987me}, recovering tree-level amplitudes by stitching together lower-point building blocks with one leg off-shell, or crossing symmetry, revealing that different processes are related through analytic continuation. From this perspective, further rethinking kinematic constraints and symmetries has led to the discovery of fascinating internal mathematical beauty of scattering amplitudes. Methods such as the use of spinor-helicity variables, improved on-shell recursion relations including BCFW~\cite{Britto:2004ap, Britto:2005fq}, or unitarity methods and generalised cuts~\cite{Bern:1994zx, Bern:1994cg, Bern:2011qt} not only deepen our understanding of quantum field theory but also facilitate the computation of complicated physical processes, as evidenced by recent achievements in high-loop calculations (see~\cite{Benincasa:2013faa, Elvang:2013cua, Weinzierl:2016bus, Cheung:2017pzi} for nice reviews).

\vskip 4pt
In the context of primordial cosmology, the foundations of cosmological correlation functions have not yet achieved the same degree of understanding. Pushing forward the state-of-the-art of computational techniques, even at first order in perturbation theory, demands gruelling mathematical dexterity, for at least three interrelated reasons. First, computing equal-time correlation functions requires integrating over the entire bulk time evolution, as time translation symmetry is lost and an asymptotically free future is absent. Second, the curved geometry of the cosmological background distorts the free propagation of particles. Finally, unlike tree-level scattering amplitudes, which are meromorphic functions of complex kinematics with branch cuts appearing only at the loop level, tree-level cosmological correlators already exhibit branch cuts. This crucial distinction arises from spontaneous particle production.

\vskip 4pt
Despite these challenges, several powerful techniques have recently been developed. At tree level, approaches such as bootstrapping correlators by solving boundary kinematic equations~\cite{Arkani-Hamed:2015bza, Arkani-Hamed:2018kmz, Pimentel:2022fsc, Jazayeri:2022kjy, Qin:2022fbv, Qin:2023ejc, Aoki:2024uyi} and reformulating perturbative diagrammatic rules in Mellin space~\cite{Sleight:2019mgd, Sleight:2019hfp, Sleight:2020obc, Sleight:2021plv} have proven effective. The use of partial Mellin-Barnes representations~\cite{Qin:2022lva, Qin:2022fbv, Xianyu:2023ytd}, leveraging the global dilatation symmetry of de Sitter space that replaces time translations, has further improved these methods. Additionally, numerical techniques that trace the time evolution of correlation functions have been developed to explore and carve out the space of most correlators~\cite{Werth:2023pfl, Pinol:2023oux, Werth:2024aui}. At the loop order, dispersive integrals have been used to tame loop diagrams~\cite{Xianyu:2022jwk, Liu:2024xyi} (also see~\cite{Chen:2016nrs, Chen:2016hrz, Chen:2018xck, Wang:2021qez, Heckelbacher:2022hbq, Qin:2023bjk, Qin:2023nhv} for other techniques).\footnote{The resummation of massless loops in de Sitter space---and the associated secular infrared divergences---has a rich and extensive story that goes beyond our focus in this work.} 

\vskip 4pt
At the same time, a deeper understanding of the structure of cosmological correlators is gradually emerging as insights from flat-space scattering amplitudes are imported to cosmology~\cite{Baumann:2022jpr}. Essentially, flat-space amplitudes are embedded within cosmological correlators as residues on the total energy pole, when energies of
the external particles add up to zero~\cite{Maldacena:2011nz, Raju:2012zr}. Similarly, on partial energy poles, when the energy of a subdiagram vanishes, correlators factorise into a product of a (shifted) lower-point correlator and a lower-point scattering amplitude. Notably, these singularities can only be reached by analytically continuing (some) external energies to negative values. Perturbative unitarity leads to a set of cutting rules~\cite{Goodhew:2020hob, Cespedes:2020xqq, Baumann:2021fxj, Melville:2021lst, Goodhew:2021oqg}, while its non-perturbative counterpart is captured through the Källén–Lehmann representation in de Sitter space~\cite{Sleight:2020obc, Hogervorst:2021uvp, Loparco:2023rug}. This opens up new possibilities to sew together simple building blocks to construct more complex correlators. Furthermore, simple cosmological correlators can be conceptualised as canonical forms of a polytope~\cite{Arkani-Hamed:2017fdk, Arkani-Hamed:2018bjr, Benincasa:2018ssx, Benincasa:2019vqr, Benincasa:2020aoj, Benincasa:2021qcb, Benincasa:2024leu, Benincasa:2024lxe} and progress has been made in understanding the differential equations they satisfy, from which ``time emerges"~\cite{Hillman:2019wgh, De:2023xue, Arkani-Hamed:2023kig, Fan:2024iek} (with related aspects discussed in~\cite{Gomez:2021qfd, Gomez:2021ujt, Armstrong:2022csc}). As we will argue in this paper, employing an \textit{off-shell} formulation of cosmological correlators not only makes their analytic and factorisation properties more explicit, but also streamlines practical computational techniques.

\vskip 4pt
Most of the analytic properties and the encoding of physical principles are phrased at the level of wavefunction coefficients, which loosely are the cosmological counterparts of scattering amplitudes.\footnote{Still, wavefunction coefficients are sensitive to total derivatives in the action and are not invariant under field redefinitions. To cure this, an $S$-matrix for de Sitter space has been recently proposed in~\cite{Melville:2023kgd}.} Yet, it remains unclear how these properties can be translated to actual observables, at least beyond leading orders in perturbation theory. Interestingly, cosmological correlators, when pushed to higher order in perturbation theory, are structurally simpler than wavefunction coefficients. For example, they share the same transcendentality as their corresponding flat-space amplitudes~\cite{Chowdhury:2023arc}. Additionally, these objects are most directly connected to cosmological data as these are the main observables. As such, they can be viewed as the cosmological analogue of cross sections.

\vskip 4pt
In this paper, we propose a systematic off-shell study of cosmological correlation functions. Our approach relies on a spectral representation of massive bulk-to-bulk propagators in de Sitter, as first introduced in~\cite{Melville:2024ove} (with previous spectral representations in de Sitter proposed in~\cite{Sleight:2020obc, Marolf:2012kh, DiPietro:2021sjt, DiPietro:2023inn}).\footnote{Integral representations of cosmological propagators have been proposed in e.g.~\cite{Meltzer:2021zin, Lee:2023kno, Donath:2024utn}, mostly for the wavefunction coefficients. But all these simple representations do not capture particle production as they were derived either for flat-space objects or for the special case of a conformally coupled field in de Sitter.} Essentially, time ordering is replaced by an appropriate contour integral. While spontaneous particle production makes it impossible to use the usual complex energy plane because of branch cuts, the key insight is instead to shift to the complex mass domain, where the dispersive integral takes the form of a split representation of off-shell bulk-to-boundary propagators. In this plane, mode functions are analytic and the integration contour can be safely closed. This approach uses a suitable $i\epsilon$ prescription to account for particle production. By incorporating particle production, unlike the usual Feynman propagator, the number of poles effectively doubles, with different Boltzmann weights that either suppress or enhance their effects. This encodes a crossing symmetry between incoming and outgoing modes separated by a Stokes line, fully capturing particle production. In the flat-space limit, where particle production is turned off, this propagator naturally reduces to the standard Feynman propagator. At the level of correlators, this off-shell propagator trivialises time integrals as they only appear in a factorised manner. 

\vskip 4pt
On a practical level, we explicitly compute the spectral integral for the four-point function of conformally coupled scalars mediated by the tree-level exchange of massive scalars, and introduce a new, partially resummed representation for this correlator. This new approach also offers a more intuitive understanding of the underlying physics. For example, the single tower of poles that we find maps directly to the quasi-normal modes of the exchanged massive field. Fundamentally, as we will show, this dispersive representation of cosmological correlators clearly illustrates the analytic structure of individual contributions and their factorisation properties. It makes explicit that exchange correlators are constructed by gluing lower-point off-shell correlators with a sewing kernel that encodes both particle production (from the off-shell leg) and a tower of EFT contributions that emerge after integrating out the exchanged field. By explicitly performing this spectral integral, we set the exchanged particle on-shell, allowing us to reconstruct more complex diagrams.

\vskip 4pt
Finally, we leverage the fundamental properties of in-in Schwinger-Keldysh diagrammatics to derive the largest-time equation satisfied by cosmological correlators. This equation relates processes occurring on different in-in branches through the analytic continuation of external energies, mirroring crossing symmetry in flat space. This relationship is universally applicable to all cosmological correlators, regardless of the order in perturbation theory or whether the theory is unitary, and it can serve as a consistency check when computing complicated diagrams. We conclude by illustrating how the largest-time equation is satisfied through a series of explicit examples.

\paragraph{Outline.} The outline of the paper is as follows: In Sec.~\ref{sec: From Time Ordering to Contour Integrals}, we briefly review the Schwinger-Keldysh formalism and cosmological propagators, with a particular focus on encoding time ordering as a contour integral. We then introduce the spectral representation of massive propagators in de Sitter and discuss their properties. In Sec.~\ref{sec: Cosmological Largest-Time Equation}, we derive the cosmological largest-time equation and prove it for both tree-level and loop-level graphs. For simple correlators, we show how their analytic structure, factorisation properties, and consistency requirements like the largest-time equation become clear when viewed from an off-shell perspective. In Sec.~\ref{sec: Massive Exchange Correlator in de Sitter}, we explicitly perform the spectral integral to obtain a new representation of a massive exchange correlator in de Sitter. We summarise our conclusions in Sec.~\ref{sec: Conclusions}.

\paragraph{Notation and Conventions.} We use the mostly-plus signature for the metric $(-, +, +, +)$. Spatial three-dimensional vectors are written in boldface $\bm{k}$. We use \textsf{sans serif} letters $(\sf{a, b, \ldots})$ for Schwinger-Keldysh indices. External and internal energies of a graph are denoted by $E_i$ and $K_i$, respectively. Cosmic (physical) time is denoted by $t$ and conformal time, such that $\mathrm{d}\tau = \mathrm{d}t/a$, by $\tau$. Throughout most of the paper, we work in the Poincaré patch of de Sitter space with the metric $\d s^2 = a^2(\tau)(-\d\tau^2+\d \bm{x}^2)$, where $a(\tau) = -(H\tau)^{-1}$ is the scale factor and $H$ is the Hubble parameter. We will often set the Hubble scale to unity $H=1$. Additional definitions will be introduced as needed in the main text.

\newpage
\section{From Time Ordering to Contour Integrals}
\label{sec: From Time Ordering to Contour Integrals}

We begin by introducing Schwinger-Keldysh propagators, which play a crucial role in calculating cosmological correlators. We then present the associated spectral representations, which encode time ordering through contour integrals in the complex plane.

\subsection{Cosmological propagators}
\label{subsec: Cosmological Correlator Diagrammatics}

We are interested in connected equal-time correlation functions of a bulk scalar field $\varphi(t, \bm{x})$. The field $\varphi$ should be thought of as the fluctuation around a classical background, as usual in cosmological settings. In the following, we consider a weakly coupled field theory described by a Lagrangian density $\mathcal{L}[\varphi]$.

\paragraph{Generating functional.} Following the standard Schwinger-Keldysh path integral formalism to compute correlation functions~\cite{Weinberg:2005vy}, two copies $\varphi_\pm$ of the field $\varphi$ are introduced on the two branches of the in-in, with the $(+)$ branch representing the forward evolution from an initial time $t_0$ to $t$ and the $(-)$ branch the backward evolution from $t$ to $t_0$. The condition that both branches are sewn at the time $t$ imposes that the field configurations $\varphi_\pm$ should coincide at $t$, i.e.~$\varphi_+(t, \bm{x}) = \varphi_-(t, \bm{x})$. Equal-time correlation functions $\braket{\Omega|\hat{\varphi}_{{\sf{a}}_1}(t, \bm{x}_1) \ldots \hat{\varphi}_{{\sf{a}}_n}(t, \bm{x}_n)|\Omega}$, with ${\sf{a}}_1, \ldots, {\sf{a}}_n=\pm$ and where $\ket{\Omega}$ is the vacuum of the fully interacting theory, are computed by taking functional derivatives 
\begin{equation}
    \braket{\Omega|\hat{\varphi}_{{\sf{a}}_1}(t, \bm{x}_1) \ldots \hat{\varphi}_{{\sf{a}}_n}(t, \bm{x}_n)|\Omega} = \frac{1}{i{{\sf{a}}_1} \ldots i{{\sf{a}}_n}} \, \frac{\delta^n Z[J_+, J_-]}{\delta J_{{\sf{a}}_1}(t, \bm{x}_1) \ldots \delta J_{{\sf{a}}_n}(t, \bm{x}_n)}\biggr\rvert_{J_\pm=0}\,,
\end{equation}
where the generating function $Z[J_+, J_-]$ is defined as
\begin{equation}
\label{eq: SK generating functional}
    Z[J_+, J_-] = \int_{\mathcal{C}_{i\epsilon}} \D \varphi_\pm \, \exp\left(i\int_{-\infty}^t\d t' \d^3x \left(\mathcal{L}[\varphi_+] + J_+ \varphi_+ - \mathcal{L}[\varphi_-] - J_- \varphi_-\right)\right) \,.
\end{equation}
The contour $\C_{i\epsilon}$ denotes the usual in-in contour that goes in the upper-half complex plane from $-\infty^+$ to $t$ and reverts towards $-\infty^-$ in the lower-half complex plane, with $-\infty^\pm \equiv -\infty(1\mp i\epsilon)$ and $\epsilon>0$ an infinitesimal parameter. The vacuum can only be defined in the asymptotic past when fluctuations have wavelengths (in Fourier space) much smaller than the size of the cosmological horizon. As such, in Eq.~(\ref{eq: SK generating functional}), the contour encodes the standard $i\epsilon$ prescription to which we will come back later.

\paragraph{Free propagators.} Resorting to perturbation theory, the full Lagrangian is split into a free part $\mathcal{L}_0$ that can be solved exactly, defining the free propagators, and an interacting part $\mathcal{L}_{\text{int}}$, i.e.~$\mathcal{L} = \mathcal{L}_0 + \mathcal{L}_{\text{int}}$. The first simplest object that can be defined is the free Wightman function $\braket{0|\hat{\varphi}(t_1, \bm{x}_1) \hat{\varphi}(t_2, \bm{x}_1)|0}$ where the expectation value is over the vacuum of the free theory $\ket{0}$. In Fourier space for the spatial coordinates only, after expanding the operator $\hat{\varphi}_{\bm{k}}(t)$ in terms of annihilation and creation operators 
\begin{equation}
    \hat{\varphi}_{\bm{k}}(t) = u_k(t) \, \hat{a}_{\bm{k}} + u_k^*(t)\,  \hat{a}^\dagger_{-\bm{k}}\,,
\end{equation}
where $u_k(t)$ is the mode function, the Wightman function is given by
\begin{equation}
\label{eq: free Wightman function}
    \G(k; t_1, t_2) = u_k(t_1) u_k^*(t_2)\,.
\end{equation}
We have used invariance under spatial translations and rotations for a homogeneous and isotropic background so that the function $\G$ only depends on a single 3-momentum magnitude $k \equiv |\bm{k}|$. Although this object does not itself have a natural physical interpretation, it is the building block for the four types of Schwinger-Keldysh propagators $\G_{\sf{ab}}(k; t_1, t_2)$ where ${\sf{a, b}} = \pm$, defined as
\begin{equation}
\label{eq: SK propagators}
    \begin{aligned}
        \G_{++}(k; t_1, t_2) &= \G(k; t_1, t_2) \Theta(t_1-t_2) + \G^*(k; t_1, t_2) \Theta(t_2-t_1)\,,\\
        \G_{+-}(k; t_1, t_2) &= \G^*(k; t_1, t_2)\,,\\
        \G_{-+}(k; t_1, t_2) &= \G(k; t_1, t_2)\,, \\
        \G_{--}(k; t_1, t_2) &= \G^*(k; t_1, t_2) \Theta(t_1-t_2) + \G(k; t_1, t_2) \Theta(t_2-t_1)\,.
    \end{aligned}
\end{equation}
These propagators being not independent is a consequence of the Wightman function satisfying the property $\G(k; t_2, t_1) = \G^*(k; t_1, t_2)$, as can be seen from Eq.~(\ref{eq: free Wightman function}). From these ``bulk-to-bulk" propagators, one can define ``bulk-to-boundary" propagators by sending $t_2\to t$ (for $t_1<t$) and using the condition that the two in-in branches coincide at the external time $\varphi_+(t, \bm{x}) = \varphi_-(t, \bm{x})$. They read
\begin{equation}
    \K_+(k; t_1) =  \G^*(k; t_1, t) \,, \quad \K_-(k; t_1) = \G(k; t_1, t) \,.
\end{equation}
Following the diagrammatic rules exposed in~\cite{Chen:2017ryl}, we represent these propagators with a black dot $\raisebox{0pt}{
\begin{tikzpicture}[line width=1. pt, scale=2]
\draw[fill=black] (0, 0) circle (.05cm);
\end{tikzpicture} 
}$ denoting $+$, a white dot $\raisebox{0pt}{
\begin{tikzpicture}[line width=1. pt, scale=2]
\draw[fill=white] (0, 0) circle (.05cm);
\end{tikzpicture} 
}$ denoting $-$, and a square $\raisebox{0pt}{
\begin{tikzpicture}[line width=1. pt, scale=2]
\draw[fill=white] ([xshift=0pt,yshift=-1.5pt]1, 0) rectangle ++(3pt,3pt);
\end{tikzpicture} 
}$ denoting the boundary at external time $t$ where $+$ and $-$ are indistinguishable:
\begin{equation}
    \begin{aligned}[c]
        \raisebox{0pt}{
\begin{tikzpicture}[line width=1. pt, scale=2]
\draw[fill=black] (0, 0) circle (.05cm) node[above=0.5mm] {$t_1$};
\draw[black] (0.05, 0) -- (1, 0);
\draw[fill=black] (1, 0) circle (.05cm) node[above=0.5mm] {$t_2$};
\end{tikzpicture} 
} &= \,\,\, \G_{++}(k; t_1, t_2)\,, \\
    \raisebox{0pt}{
\begin{tikzpicture}[line width=1. pt, scale=2]
\draw[fill=black] (0, 0) circle (.05cm) node[above=0.5mm] {$t_1$};
\draw[black] (0.05, 0) -- (1, 0);
\draw[fill=white] (1, 0) circle (.05cm) node[above=0.5mm] {$t_2$};
\end{tikzpicture} 
} &= \,\,\, \G_{+-}(k; t_1, t_2)\,, \\
    \raisebox{0pt}{
\begin{tikzpicture}[line width=1. pt, scale=2]
\draw[fill=white] (0, 0) circle (.05cm) node[above=0.5mm] {$t_1$};
\draw[black] (0.05, 0) -- (1, 0);
\draw[fill=black] (1, 0) circle (.05cm) node[above=0.5mm] {$t_2$};
\end{tikzpicture} 
} &= \,\,\, \G_{-+}(k; t_1, t_2)\,, \\
    \raisebox{0pt}{
\begin{tikzpicture}[line width=1. pt, scale=2]
\draw[fill=white] (0, 0) circle (.05cm) node[above=0.5mm] {$t_1$};
\draw[black] (0.05, 0) -- (1, 0);
\draw[fill=white] (1, 0) circle (.05cm) node[above=0.5mm] {$t_2$};
\end{tikzpicture} 
} &= \,\,\, \G_{--}(k; t_1, t_2)\,,
    \end{aligned}
    \qquad
    \begin{aligned}[c]
    \raisebox{0pt}{
\begin{tikzpicture}[line width=1. pt, scale=2]
\draw[fill=black] (0, 0) circle (.05cm) node[above=0.5mm] {$t_1$};
\draw[black] (0.05, 0) -- (1, 0);
\draw[fill=white] ([xshift=0pt,yshift=-1.5pt]1, 0) rectangle ++(3pt,3pt);
\end{tikzpicture} 
} &= \,\,\, \K_{+}(k; t_1)\,, \\
    \raisebox{0pt}{
\begin{tikzpicture}[line width=1. pt, scale=2]
\draw[fill=white] (0, 0) circle (.05cm) node[above=0.5mm] {$t_1$};
\draw[black] (0.05, 0) -- (1, 0);
\draw[fill=white] ([xshift=0pt,yshift=-1.5pt]1, 0) rectangle ++(3pt,3pt);
\end{tikzpicture} 
} &= \,\,\, \K_{-}(k; t_1)\,.
    \end{aligned}
\end{equation}
A double-coloured dot $\raisebox{0pt}{
\begin{tikzpicture}[line width=1. pt, scale=2]
\draw (0, 0) circle (.05cm);
\fill[black] (0,0) -- (90:0.05) arc (90:270:0.05) -- cycle;
\end{tikzpicture} 
}$ means either a black or a white dot. 

\subsection{Multiple propagators: a matter of contour}

The previously defined propagators are nothing but Green's functions of the free equation of motion with appropriate boundary conditions, and therefore can be written as contour integrals.

\paragraph{Feynman propagator.} To illustrate how a frequency-space representation of the propagators can be constructed and for the sake of simplicity, let us first consider a massless scalar field $\varphi$ in flat space. We will upgrade the presented integral representations of propagators to massive fields in de Sitter space in the next section. The corresponding positive-frequency mode function reads $u_k(t) = \tfrac{e^{-ik t}}{\sqrt{2k}}$. The generalisation to massive fields is straightforward after setting $k\rightarrow\omega_k=\sqrt{k^2+m^2}$. The key insight is to realise that time ordering can be written as a frequency-space integral using the mathematical identity
\begin{equation}
\label{eq: flat-space mathematical trick}
    e^{-ik(t_1 - t_2)} \Theta(t_1-t_2) + e^{+ik(t_1 - t_2)}\Theta(t_2-t_1) = -2k \int_{-\infty}^{+\infty} \frac{\d \omega}{2i\pi} \frac{e^{i\omega(t_1-t_2)}}{(\omega^2 - k^2)_{i\epsilon}}\,,
\end{equation}
where the $i\epsilon$ prescription is implemented as follows
\begin{equation}
\label{eq: flat-space iepsilon prescription}
    \frac{1}{(\omega^2-k^2)_{i\epsilon}} \equiv \frac{1}{2k}\left[\frac{1}{\omega - (k-i\epsilon)} - \frac{1}{\omega - (-k+i\epsilon)}\right] = \frac{1}{\omega^2-k^2+i\epsilon}\,,
\end{equation}
with $\epsilon>0$ an infinitesimal positive parameter, and where in the second equality we have dropped terms of order $\mathcal{O}(\epsilon^2)$ and set $2k\epsilon = \epsilon$, which is valid in the limit $\epsilon \to 0$. This procedure effectively sets the two poles slightly off the real axis in the complex plane. To see that we indeed recover the correct Feynman propagator, we suppose that $t_1>t_2$ and close the contour in the upper-half complex plane. The first term in Eq.~(\ref{eq: flat-space iepsilon prescription}) gives zero and only the second term selects the residue, effectively setting the exponential on shell, i.e.~$e^{i\omega(t_1-t_2)} \to e^{-ik(t_1-t_2)}$. One can proceed in the same way for $t_1<t_2$, this way closing the contour in the lower-half complex plane, to find $e^{ik(t_1-t_2)}$. Importantly, this requires that the mode function should be analytic, which is valid here for the exponential in the entire complex plane, and that the integrand decays at least faster than $1/\omega$ when $\omega \to \infty$ so that the semi-circle at infinity does not contribute.

\vskip 4pt
In the end, the time-ordered propagator $\G_{++}$ can be re-written as
\begin{equation}
\label{eq: flat-space bulk-to-bulk propagator}
    \G_{++}(k; t_1, t_2) = i\,\int_{-\infty}^{+\infty} \frac{\d\omega}{2\pi} \frac{\tilde{u}^*_\omega(t_1) \tilde{u}_\omega(t_2)}{(\omega^2 - k^2)_{i\epsilon}}\,,
\end{equation}
where $\tilde{u}_k(t) = e^{-ikt}$ is the mode function after stripping off the overall normalisation. Naturally, we have $\G_{--}(k; t_1, t_2) = \G_{++}^*(k; t_1, t_2)$. Some comments are in order. First, the use of tilted mode functions $\tilde{u}_k(t)$ avoids introducing unnecessary branch cuts in the integrand coming from $\sqrt{k}$, thereby preserving its analytic property. Second, both times $t_1$ and $t_2$ appear factorised so that time integrals within correlators become trivial. This property will be illustrated with concrete examples in Sec.~\ref{subsec: application to simple diagrams}. Finally, the $i\epsilon$ prescription is just a simple way for representing time ordering and specifying a pole prescription. Equivalently, one may deform the contour slightly from $-\infty^- = -\infty(1+i\epsilon)$ to $+\infty^+ = +\infty(1+i\epsilon)$, while keeping the poles on the real axis. We show in App.~\ref{eq: ieps from wavefunctional} how this $i\epsilon$ prescription can be recovered directly from the Schwinger-Keldysh path integral.

\paragraph{Retarded and advanced propagators.} Various Green's functions with different boundary conditions can be recovered from the usual integral representation of the Feynman propagator. This can be easily done by deforming and combining diverse integration contours. For example, using the standard identity satisfied by the Heaviside function $\Theta(t_1-t_2) + \Theta(t_2-t_1) = 1$,
the time-ordered propagator $\G_{++}$ can be written as
\begin{equation}
\label{eq: ++ propagator in terms of +- and retarded}
    \G_{++}(k; t_1, t_2) = \G^*(k; t_1, t_2) + \left[\G(k; t_1, t_2) - \G^*(k; t_1, t_2)\right]\Theta(t_1-t_2)\,.
\end{equation}
The first term is the non-time-ordered propagator $\G_{+-}(k; t_1, t_2)$ and therefore is factorised in time, whereas the second one is nothing but the causal retarded Green's function $\G_R(k; t_1, t_2)$. In terms of integration contours, Eq.~(\ref{eq: ++ propagator in terms of +- and retarded}) reads
\begin{equation}
        \vcenter{\hbox{
\begin{tikzpicture}[line width=1. pt, scale=2]
\draw[black, -] (-0.5,0) -- (0.5,0) coordinate (xaxis);
\draw[black, -] (0,-0.4) -- (0,0.4) coordinate (yaxis);
\draw[black, fill = black] (0.2, -0.1) circle (.02cm);
\draw[black, fill = black] (-0.2, 0.1) circle (.02cm);
\path[pyred, draw, line width = 1pt, postaction = decorate, decoration={markings,
			mark=at position 0.7 with {\arrow[line width=1pt]{>}}}] (-0.4, 0) -- (0.4, 0);
\end{tikzpicture}}}
=
        \vcenter{\hbox{
\begin{tikzpicture}[line width=1. pt, scale=2]
\draw[black, -] (-0.5,0) -- (0.5,0) coordinate (xaxis);
\draw[black, -] (0,-0.4) -- (0,0.4) coordinate (yaxis);
\draw[black, fill = black] (0.2, 0) circle (.02cm);
\draw[black, fill = black] (-0.2, 0) circle (.02cm);
\path[pyred, draw, line width = 1pt, postaction = decorate, decoration={markings,
			mark=at position 0.4 with {\arrow[line width=1pt]{>}}}] (0.3, 0) arc (0:-360:0.1) -- (0.3, 0);
\end{tikzpicture}}}
\hspace*{0.22cm} - 
\vcenter{\hbox{
\begin{tikzpicture}[line width=1. pt, scale=2]
\draw[black, -] (-0.5,0) -- (0.5,0) coordinate (xaxis);
\draw[black, -] (0,-0.4) -- (0,0.4) coordinate (yaxis);
\draw[black, fill = black] (0.2, 0) circle (.02cm);
\draw[black, fill = black] (-0.2, 0) circle (.02cm);
\path[pyred, draw, line width = 1pt, postaction = decorate, decoration={markings,
			mark=at position 0.7 with {\arrow[line width=1pt]{>}}}] (-0.4, -0.2) -- (0.4, -0.2);
\end{tikzpicture}}}\,.
\end{equation}
Importantly, non-local processes lead to a vanishing retarded Greens's function. This observation has consequences for the cosmological case, and enables to extract various signals from correlators with different origin. Similarly, one can flip the second Heaviside function in the expression of $\G_{++}$ to obtain
\begin{equation}
    \G_{++}(k; t_1, t_2) = \G(k; t_1, t_2) + \left[\G^*(k; t_1, t_2) - \G(k; t_1, t_2)\right]\Theta(t_2-t_1)\,,
\end{equation}
which, in terms of integration contours, reads
\begin{equation}
        \vcenter{\hbox{
\begin{tikzpicture}[line width=1. pt, scale=2]
\draw[black, -] (-0.5,0) -- (0.5,0) coordinate (xaxis);
\draw[black, -] (0,-0.4) -- (0,0.4) coordinate (yaxis);
\draw[black, fill = black] (0.2, -0.1) circle (.02cm);
\draw[black, fill = black] (-0.2, 0.1) circle (.02cm);
\path[pyred, draw, line width = 1pt, postaction = decorate, decoration={markings,
			mark=at position 0.7 with {\arrow[line width=1pt]{>}}}] (-0.4, 0) -- (0.4, 0);
\end{tikzpicture}}}
=
        \vcenter{\hbox{
\begin{tikzpicture}[line width=1. pt, scale=2]
\draw[black, -] (-0.5,0) -- (0.5,0) coordinate (xaxis);
\draw[black, -] (0,-0.4) -- (0,0.4) coordinate (yaxis);
\draw[black, fill = black] (0.2, 0) circle (.02cm);
\draw[black, fill = black] (-0.2, 0) circle (.02cm);
\path[pyred, draw, line width = 1pt, postaction = decorate, decoration={markings,
			mark=at position 0.4 with {\arrow[line width=1pt]{>}}}] (-0.1, 0) arc (0:360:0.1) -- (-0.1, 0);
\end{tikzpicture}}}
\hspace*{0.22cm} - 
\vcenter{\hbox{
\begin{tikzpicture}[line width=1. pt, scale=2]
\draw[black, -] (-0.5,0) -- (0.5,0) coordinate (xaxis);
\draw[black, -] (0,-0.4) -- (0,0.4) coordinate (yaxis);
\draw[black, fill = black] (0.2, 0) circle (.02cm);
\draw[black, fill = black] (-0.2, 0) circle (.02cm);
\path[pyred, draw, line width = 1pt, postaction = decorate, decoration={markings,
			mark=at position 0.7 with {\arrow[line width=1pt]{>}}}] (-0.4, 0.2) -- (0.4, 0.2);
\end{tikzpicture}}}\,.
\end{equation}

\subsection{Spectral representation of massive cosmological propagators}
\label{subsec: Spectral representation of massive cosmological propagators}

We now show how the previous integration contour prescription is modified when taking into account spontaneous particle production. For concreteness in what follows, we consider a real massive scalar field in de Sitter, see~App.~\ref{sec: More on massive fields in de Sitter} for more details.\footnote{We consider fields in the principal series $m/H\geq 3/2$ so that $\mu \equiv \sqrt{m^2/H^2 - 9/4}$ is real. In particular, we will not deal with late-time secular divergences appearing when exchanging light fields.}

\paragraph{Contour prescription.} Spontaneous particle production manifests itself as non-analyticity in the complex energy domain $k\to \omega$ once the momentum is set off shell. This renders impossible to construct an integral representation of massive cosmological propagators for the reason that implementing time ordering requires crossing the branch cut. This challenge can be alleviated by instead analytically continuing the massive mode function to the complex $\mu$ plane, allowing the field to acquire any complex mass, as shown in~\cite{Melville:2024ove}. The integral representation of the time-ordered massive cosmological propagator reads\footnote{The spectral parameter $\nu$ should not be confused with $\nu = i\mu$ that is commonly used to label light fields in the complementary series.}
\begin{equation}
\label{eq: massive propagator in de Sitter}
    \G_{++}(k; \tau_1, \tau_2) = i\,\int_{-\infty}^{+\infty} \d \nu \, \N_\nu \, \frac{u_k^*(\tau_1, \nu) u_k^*(\tau_2, \nu)}{(\nu^2 - \mu^2)_{i\epsilon}}\,,
\end{equation}  
where $\N_\nu \equiv \tfrac{\nu}{\pi}\sinh(\pi\nu)$ is the de Sitter density of states, $u_k^*(\tau, \mu)$ is the negative-frequency massive mode function defined in~(\ref{eq: massive dS mode function neg}), and the contour prescription to select the correct pole structure is given by\footnote{The explicit pole structure (and the corresponding residues) can be immediately obtained from the usual flat-space prescription~(\ref{eq: flat-space iepsilon prescription}) or by noticing that
\begin{equation}
    \frac{1}{\nu^2 - \mu^2 + i\epsilon} = \frac{\Gamma(i\nu-i\mu-\epsilon)\Gamma(-i\nu-i\mu-\epsilon)}{\Gamma(i\nu-i\mu-\epsilon+1)\Gamma(-i\nu-i\mu-\epsilon+1)}\,.
\end{equation}}
\begin{equation}
\label{eq: de Sitter iepsilon prescription}
    \begin{aligned}
        \frac{1}{(\nu^2-\mu^2)_{i\epsilon}} \equiv \frac{1}{2\sinh(\pi\mu)} \left[\frac{e^{+\pi\mu}}{\nu^2-\mu^2+i\epsilon} - \frac{e^{-\pi\mu}}{\nu^2-\mu^2-i\epsilon}\right]\,.
    \end{aligned}
\end{equation}
Naturally, the anti-time ordered propagator is given by $\G_{--}(k; \tau_1, \tau_2) = \G_{++}^*(k; \tau_1, \tau_2)$. The detailed derivation of~(\ref{eq: massive propagator in de Sitter}) is presented in App.~\ref{subsec: Derivation of the spectral representation}. The integral over the spectral parameter $\nu$ can be viewed as an integration over the scaling dimension $\Delta = \tfrac{3}{2}+i\nu$ in the principal series that labels infinite-dimensional (scalar) representations of the de Sitter group, which defines eigenvalues of the quadratic Casimir operator $\Delta (3-\Delta)$, see e.g.~\cite{Basile:2016aen, Karateev:2018oml, Sengor:2019mbz, Sun:2021thf}. In this sense, this integral is closely related to the de Sitter Källén–Lehmann representation albeit in spatial Fourier space~\cite{Hogervorst:2021uvp, Loparco:2023rug}. Such spectral representation has its analogue in anti-de Sitter space, see e.g.~\cite{Penedones:2010ue}, and in other curved geometries~\cite{Moschella:2007zza}.

\paragraph{Flat-space limit.} The spectral representation~(\ref{eq: massive propagator in de Sitter}) extends the familiar flat-space prescription~(\ref{eq: flat-space iepsilon prescription}) by capturing spontaneous particle production while at the same time having a well-defined flat-space limit. Indeed, in the limit $H\to0$, the Boltzmann suppression is turned off $e^{-\pi\mu}\to0$ as $\mu\sim m/H$ for a fixed mass. The additional unusual poles $\nu = \pm \mu \pm i\epsilon$ vanish and the contour prescription reduces to the flat-space one~(\ref{eq: flat-space iepsilon prescription})
\begin{equation}
    \frac{1}{(\nu^2-\mu^2)_{i\epsilon}} \xrightarrow[\mu \to \infty]{} \frac{1}{\nu^2-\mu^2+i\epsilon}\,.
\end{equation}
Following the derivation of~(\ref{eq: massive propagator in de Sitter}) presented in App.~\ref{subsec: Derivation of the spectral representation}, for $\tau_1$ being in the future of $\tau_2$, and using limiting forms given in App.~\ref{subsec: Useful formulae}, the massive propagator reduces to
\begin{equation}
    \begin{aligned}
    \G_{++}(k; \tau_1, \tau_2) &= \frac{i}{2}\int_{-\infty}^{+\infty} \d\nu \, \frac{\nu J_{i\nu}(-k\tau_1) H_{i\nu}^{(2)}(-k\tau_2)}{\nu^2-\mu^2+i\epsilon} = \frac{\pi}{2} J_{i\mu}(-k\tau_1) H_{i\mu}^{(2)}(-k\tau_2) \\
    &\xrightarrow[\mu \to \infty]{} \frac{e^{-i\mu(t_1-t_2)}}{\sqrt{2\mu}}\,,
    \end{aligned}
\end{equation}
where we have selected the \textit{single} pole at $\nu = \mu-i\epsilon$ by closing the contour in the lower-half complex plane, and have set $H=1$. We recover the properly-normalised flat-space propagator composed of plane waves oscillating in cosmic time $t$. In the flat-space limit, this propagator captures the free super-horizon evolution of the massive field when spatial gradients are negligible, leading to a dispersion relation dominated by the mass $\omega_k \sim \mu$. This provides insight into why analytically continuing the mass, rather than the momentum, allows for defining a spectral representation of massive fields.

\paragraph{Particle production.} Particle production of massive fields manifests itself as the emergence of negative-frequency modes over time, when initialised with a positive-frequency mode as selected by the Bunch-Davies vacuum. This process leads to mode mixing, reflected in a non-zero average particle number observed at later times. The Bessel function defines the mode function in the infinite future when the massive field reaches a steady state, as
\begin{equation}
    v_k(\tau, \mu) = \frac{\Gamma(1+i\mu)}{\sqrt{2\mu}} \left(\frac{k}{2}\right)^{-i\mu} J_{i\mu}(-k\tau) \xrightarrow[\tau \to 0]{} \frac{e^{-i\mu t}}{\sqrt{2\mu}}\,.
\end{equation}
Notice in passing that the late-time limit $\tau\to0$ equivalently also corresponds to the large-mass limit $\mu\to\infty$ as can easily recovered using~(\ref{eq: large-order asymptotic expansion}) and the Stirling formula $\Gamma(z) \sim e^{-z} z^z \sqrt{2\pi/z}$. This is not surprising as both variables are conjugate to each other. Decomposing this outgoing mode function onto the basis of ingoing ones, we obtain
\begin{equation}
    v_k(\tau, \mu) = \alpha_k u_k(\tau, \mu) + \beta_k u_k^*(\tau, \mu)\,,
\end{equation}
where the Bogolyubov coefficients are
\begin{equation}
    \alpha_k = \frac{\Gamma(1+i\mu)}{\sqrt{2\pi\mu}} \left(\frac{k}{2}\right)^{-i\mu} e^{+\frac{\pi\mu}{2}-\frac{i\pi}{2}}\,, \quad \beta_k = \frac{\Gamma(1+i\mu)}{\sqrt{2\pi\mu}} \left(\frac{k}{2}\right)^{-i\mu} e^{-\frac{\pi\mu}{2}+\frac{i\pi}{2}}\,,
\end{equation}
from which we obtain the mean particle density of ``out" excitations with wavenumber 
$\bm{k}$ in the ``in" vacuum state  $|\beta_k|^2 = 1/(e^{2\pi\mu}-1)$. We recover the usual Bose-Einstein distribution, when identifying the energy with the rest mass $\mu$ in a thermal bath set by the de Sitter temperature $T_{\text{dS}} = 1/2\pi$.

\begin{figure}[t!]
   \hspace*{0.5cm}
    \includegraphics[width=0.5\textwidth]{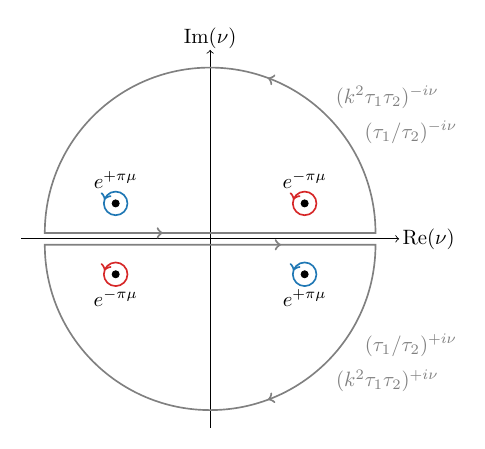}
   \caption{Illustration of the contour prescription~(\ref{eq: de Sitter iepsilon prescription}) in the complex $\nu$ plane. The usual poles \textcolor{pyblue}{$\nu = \pm \mu \mp i\epsilon$} effectively project the off-shell mode onto late-time positive-frequency modes, whereas taking the residues at the Boltzmann suppressed poles \textcolor{pyred}{$\nu = \pm \mu \pm i\epsilon$} projects the off-shell mode onto the late-time negative-frequency modes. This interchange of ingoing and outgoing modes captures particle production. In the late-time limit, the $\nu \leftrightarrow-\nu$ symmetry of the off-shell modes completely fixes the contour prescription both for the local and non-local contributions, leading to the disappearance of time ordering.}
  \label{fig: de Sitter integration contour prescription}
\end{figure}

\paragraph{Local and non-local parts.} The $i\epsilon$ prescription~(\ref{eq: de Sitter iepsilon prescription}) projects ingoing particle states onto outgoing states, which enables the mixing of positive- and negative-frequency mode functions. This is made manifest by rewriting the off-shell ingoing mode function in terms of off-shell outgoing ones using the connection formula~(\ref{eq: connection formula}), so that the propagator can be written~\cite{Arkani-Hamed:2015bza, Lee:2016vti, Tong:2021wai}
\begin{equation}
    \G_{++}(k; \tau_1, \tau_2) = \G_{++}^{\text{local}}(k; \tau_1, \tau_2) + \G_{++}^{\text{non-local}}(k; \tau_1, \tau_2)\,,
\end{equation}
where both contributions read
\begin{equation}
    \begin{aligned}
        \G_{++}^{\text{local}}(k; \tau_1, \tau_2) &= \frac{+i\pi}{4} \int_{-\infty}^{+\infty} \d\nu \,\frac{\N_\nu}{\sinh^2(\pi\nu)}\, \frac{J_{i\nu}(-k\tau_1) J_{i\nu}^*(-k\tau_2) + (\nu \leftrightarrow-\nu)}{(\nu^2-\mu^2)_{i\epsilon}} \,, \\
        \G_{++}^{\text{non-local}}(k; \tau_1, \tau_2) &= \frac{-i\pi}{4} \int_{-\infty}^{+\infty} \d\nu \,\frac{\N_\nu}{\sinh^2(\pi\nu)}\, \frac{e^{-\pi\nu}J_{i\nu}(-k\tau_1) J_{i\nu}(-k\tau_2) + (\nu\leftrightarrow-\nu)}{(\nu^2-\mu^2)_{i\epsilon}} \,.
    \end{aligned}
\end{equation}
The local part is interpreted as the accumulated dynamical phase of a single particle propagating since in the late-time limit we have
\begin{equation}
    J_{i\nu}(-k\tau_1) J_{i\nu}^*(-k\tau_2) \sim \frac{\sinh(\pi\nu)}{\pi\nu} \left(\frac{\tau_1}{\tau_2}\right)^{i\nu}\propto e^{-i\nu(t_1-t_2)}\,.
\end{equation}
Notably, the local part is independent of the momentum so that it describes correlation at coincident points in position space. Assuming $\tau_1>\tau_2$, we close the contour in the lower-half complex plane for the term $J_{i\nu}(-k\tau_1) J_{i\nu}^*(-k\tau_2)$ and in the upper-half complex plane for the complex conjugate piece, which leads to
\begin{equation}
    \begin{aligned}
        \G_{++}^{\text{local}}(k; \tau_1, \tau_2) &= \frac{\pi}{4\sinh^2(\pi\mu)} \left[e^{+\pi\mu} J_{i\mu}(-k\tau_1)J_{i\mu}^*(-k\tau_2) + e^{-\pi\mu} J_{i\mu}^*(-k\tau_1)J_{i\mu}(-k\tau_2)\right] \\
        &\xrightarrow[\tau_1, \tau_2 \to 0]{} \frac{\Gamma(+i\mu)\Gamma(-i\mu)}{4\pi} \left[e^{+\pi\mu} \left(\frac{\tau_1}{\tau_2}\right)^{+i\mu} + e^{-\pi\mu} \left(\frac{\tau_1}{\tau_2}\right)^{-i\mu}\right]\,.
    \end{aligned}
\end{equation}
For the reverse time ordering, the contours should be closed in the opposite directions for both contributions. The doubling of poles projects the off-shell positive-frequency mode into positive- and negative-frequency ones as $(\tfrac{\tau_1}{\tau_2})^{\pm i\nu} \to e^{\pm\pi\mu}(\tfrac{\tau_1}{\tau_2})^{\pm i\mu} + e^{\mp\pi\mu}(\tfrac{\tau_1}{\tau_2})^{\mp i\mu}$, which then combine to reconstruct the correct mode mixing. This is illustrated in Fig.~\ref{fig: de Sitter integration contour prescription}. On the other hand, the non-local part corresponds to the creation and the propagation of two particles since
\begin{equation}
    J_{i\nu}(-k\tau_1) J_{i\nu}(-k\tau_2) \sim  \frac{1}{\Gamma(1+i\nu)^2} \left(\frac{k^2\tau_1\tau_2}{4}\right)^{i\nu} \propto e^{-i\nu(t_1-t_\star)} e^{-i\nu(t_2-t_\star)}\,,
\end{equation}
in the late-time limit, where $t_\star=\log k$ is the pair production (cosmic) time. Being non-analytic in the momentum, this piece describes long-range non-local correlations in position space. Similarly to the local contribution, we close the contour in the lower-half complex plane for the term $e^{-\pi\nu}J_{i\nu}(-k\tau_1) J_{i\nu}(-k\tau_2)$ and in the opposite direction for the second one. We obtain
\begin{equation}
    \begin{aligned}
        \G_{++}^{\text{non-local}}(k; \tau_1, \tau_2) &= \frac{\pi}{4\sinh^2(\pi\mu)} \left[J_{i\mu}(-k\tau_1)J_{i\mu}(-k\tau_2) + J_{i\mu}^*(-k\tau_1)J_{i\mu}^*(-k\tau_2)\right] \\
        &\xrightarrow[\tau_1, \tau_2 \to 0]{} \frac{1}{4\pi} \left[\Gamma(-i\mu)^2 \left(\frac{k\tau_1\tau_2}{4}\right)^{+i\mu} + \Gamma(+i\mu)^2 \left(\frac{k\tau_1\tau_2}{4}\right)^{-i\mu}\right]\,.
    \end{aligned}
\end{equation}
Importantly, at late times, the choice of closing the contour is completely fixed so that time ordering does not matter. The Feynman propagator $\G_{++}$ effectively reduces to the Wightman function $\G$. Non-local processes being real, as can be explicitly seen from the last equation, they do not enter the advanced or retarded Green's functions. This reflects the fact that the endpoints of a soft propagator are in space-like separation. With the spectral representation of the time-ordered propagator, subtracting the commutator of field operators simply amounts to selecting the residues at the poles in the contour prescription
\begin{equation}
    \begin{aligned}
        \G_{++}^{\text{non-local}}(k; \tau_1, \tau_2) &= \G_{++}(k; \tau_1, \tau_2) \,\, - \,\,\G_R(k; \tau_1, \tau_2)\,, \\
        \vcenter{\hbox{
\begin{tikzpicture}[line width=1. pt, scale=2]
\draw[black, -] (-0.5,0) -- (0.5,0) coordinate (xaxis);
\draw[black, -] (0,-0.4) -- (0,0.4) coordinate (yaxis);
\draw[black, fill = black] (0.2, 0.1) circle (.02cm);
\draw[black, fill = black] (0.2, -0.1) circle (.02cm);
\draw[black, fill = black] (-0.2, 0.1) circle (.02cm);
\draw[black, fill = black] (-0.2, -0.1) circle (.02cm);
\path[pyred, draw, line width = 1pt, postaction = decorate, decoration={markings,
			mark=at position 0.4 with {\arrow[line width=1pt]{>}}}] (0.3, -0.1) arc (0:-360:0.1) -- (0.3, -0.1);
\path[pyred, draw, line width = 1pt, postaction = decorate, decoration={markings,
			mark=at position 0.4 with {\arrow[line width=1pt]{>}}}] (-0.1, -0.1) arc (0:-360:0.1) -- (-0.1, -0.1);
\end{tikzpicture}}}
&=
        \vcenter{\hbox{
\begin{tikzpicture}[line width=1. pt, scale=2]
\draw[black, -] (-0.5,0) -- (0.5,0) coordinate (xaxis);
\draw[black, -] (0,-0.4) -- (0,0.4) coordinate (yaxis);
\draw[black, fill = black] (0.2, 0.1) circle (.02cm);
\draw[black, fill = black] (0.2, -0.1) circle (.02cm);
\draw[black, fill = black] (-0.2, 0.1) circle (.02cm);
\draw[black, fill = black] (-0.2, -0.1) circle (.02cm);
\path[pyred, draw, line width = 1pt, postaction = decorate, decoration={markings,
			mark=at position 0.7 with {\arrow[line width=1pt]{>}}}] (-0.4, 0) -- (0.4, 0);
\end{tikzpicture}}}
\hspace*{0.22cm} - 
\vcenter{\hbox{
\begin{tikzpicture}[line width=1. pt, scale=2]
\draw[black, -] (-0.5,0) -- (0.5,0) coordinate (xaxis);
\draw[black, -] (0,-0.4) -- (0,0.4) coordinate (yaxis);
\draw[black, fill = black] (0.2, 0.1) circle (.02cm);
\draw[black, fill = black] (0.2, -0.1) circle (.02cm);
\draw[black, fill = black] (-0.2, 0.1) circle (.02cm);
\draw[black, fill = black] (-0.2, -0.1) circle (.02cm);
\path[pyred, draw, line width = 1pt, postaction = decorate, decoration={markings,
			mark=at position 0.7 with {\arrow[line width=1pt]{>}}}] (-0.4, -0.2) -- (0.4, -0.2);
\end{tikzpicture}}}\,.
    \end{aligned}
\end{equation}
Of course, selecting the poles in upper-half complex plane requires subtracting the advanced Green's function $\G_A$. This contour deformation is completely equivalent to the bulk cutting rules that extract non-local signals~\cite{Tong:2021wai}.

\paragraph{Conformally coupled field.} The special case of a conformally coupled field in de Sitter $m^2=2H^2$ is related to the flat-space one by a conformal transformation. Physically, particle production of a conformally coupled field in de Sitter is effectively turned off so that its mode function is analytic in the entire complex $k$ plane. Therefore, from the flat-space propagator~(\ref{eq: flat-space bulk-to-bulk propagator}), one immediately obtains
\begin{equation}
    \G_{++}(k; \tau_1, \tau_2) = i\, \tau_1 \tau_2 \, \int_{-\infty}^{+\infty} \frac{\d\omega}{2\pi} \frac{\tilde{u}^*_\omega(\tau_1) \tilde{u}_\omega(\tau_2)}{(\omega^2 - k^2)_{i\epsilon}}\,,
\end{equation}
where $\tilde{u}_k(\tau) = e^{-ik\tau}$ and where we have set $H=1$.

\section{Cosmological Largest-Time Equation}
\label{sec: Cosmological Largest-Time Equation}

Unitarity at the quantum level, often described as the conservation of probability, imposes powerful constraints on observables, even when the full set of states in the Hilbert space is unknown. At a perturbative level, unitarity implies relations between cosmological correlators~\cite{Goodhew:2020hob} and conserved quantities under unitary time evolution~\cite{Cespedes:2020xqq}. These relations are analogous to the flat-space optical theorem~\cite{Weinberg:1996kr, Schwartz:2014sze}. It has been shown that these relations can be organised into a set of cutting rules~\cite{Meltzer:2020qbr, Melville:2021lst, Goodhew:2021oqg, Baumann:2021fxj, Meltzer:2021zin}, which stem from the Hermitian analytic nature of propagators and the straightforward factorisation of a combination of bulk-to-bulk propagators, bearing resemblance to the standard $S$-matrix Cutkosky rules, see e.g.~\cite{Britto:2024mna} for a recent review. However, these cutting rules were derived for the quantum mechanical wavefunction of the universe, as relations satisfied by the corresponding wavefunction coefficients.

\vskip 4pt
In this section, we derive similar relations directly at the level of cosmological correlators, and relaxing the assumption of unitarity. We show that individual in-in branch components of cosmological correlators, once external legs are properly analytically continued to negative energies on the negative branch, satisfy the largest-time equation~\cite{Veltman:1994wz}. We then showcase how this equation is satisfied for relatively simple diagrams.

\subsection{Propagator identity from an off-shell perspective}

Using the spectral representation of bulk-to-bulk propagators presented in the previous section, deriving the largest-time equation exactly parallels that for the flat-space $S$-matrix~\cite{Cutkosky:1960sp, tHooft:1973wag, Veltman:1994wz}. The essence of this equation lies in a specific trivial identity satisfied by propagators. Using the standard distributional identity
\begin{equation}
    \frac{1}{x+i\epsilon} - \frac{1}{x-i\epsilon} = -2i\pi\delta(x)\,,
\end{equation}
in the limit $\epsilon\to 0$, which follows from the Sokhotski-Plemelj identity, the flat-space bulk-to-bulk propagators~(\ref{eq: flat-space bulk-to-bulk propagator}) satisfy the known relation\footnote{In addition, we have used the following identity for the delta Dirac function
\begin{equation}
    \delta(\omega^2 - k^2) = \frac{1}{2k} [\delta(\omega+k) + \delta(\omega-k)] \,,
\end{equation}
for $k>0$.}
\begin{equation}
\label{eq: propagator identity}
    \G_{++}(k; t_1, t_2) + \G_{--}(k; t_1, t_2) = \G_{+-}(k; t_1, t_2) + \G_{-+}(k; t_1, t_2)\,,
\end{equation}
that is, since $\G_{--}(k; t_1, t_2) = \G_{++}^*(k; t_1, t_2)$, the real part does not contain time ordering. In terms of integration contour, this identity reads
\begin{equation}
\label{eq: flat-space propagator identity integration contour}
      \vcenter{\hbox{
\begin{tikzpicture}[line width=1. pt, scale=2]
\draw[black, -] (-0.5,0) -- (0.5,0) coordinate (xaxis);
\draw[black, -] (0,-0.4) -- (0,0.4) coordinate (yaxis);
\draw[black, fill = black] (0.2, -0.1) circle (.02cm);
\draw[black, fill = black] (-0.2, 0.1) circle (.02cm);
\path[pyred, draw, line width = 1pt, postaction = decorate, decoration={markings,
			mark=at position 0.7 with {\arrow[line width=1pt]{>}}}] (-0.4, 0) -- (0.4, 0);
\end{tikzpicture}}}
\hspace*{0.22cm} - 
\vcenter{\hbox{
\begin{tikzpicture}[line width=1. pt, scale=2]
\draw[black, -] (-0.5,0) -- (0.5,0) coordinate (xaxis);
\draw[black, -] (0,-0.4) -- (0,0.4) coordinate (yaxis);
\draw[black, fill = black] (0.2, 0.1) circle (.02cm);
\draw[black, fill = black] (-0.2, -0.1) circle (.02cm);
\path[pyred, draw, line width = 1pt, postaction = decorate, decoration={markings,
			mark=at position 0.7 with {\arrow[line width=1pt]{>}}}] (-0.4, 0) -- (0.4, 0);
\end{tikzpicture}}}
=
        \vcenter{\hbox{
\begin{tikzpicture}[line width=1. pt, scale=2]
\draw[black, -] (-0.5,0) -- (0.5,0) coordinate (xaxis);
\draw[black, -] (0,-0.4) -- (0,0.4) coordinate (yaxis);
\draw[black, fill = black] (0.2, 0) circle (.02cm);
\draw[black, fill = black] (-0.2, 0) circle (.02cm);
\path[pyred, draw, line width = 1pt, postaction = decorate, decoration={markings,
			mark=at position 0.4 with {\arrow[line width=1pt]{>}}}] (0.3, 0) arc (0:-360:0.1) -- (0.3, 0);
\end{tikzpicture}}}
\hspace*{0.22cm} - 
\vcenter{\hbox{
\begin{tikzpicture}[line width=1. pt, scale=2]
\draw[black, -] (-0.5,0) -- (0.5,0) coordinate (xaxis);
\draw[black, -] (0,-0.4) -- (0,0.4) coordinate (yaxis);
\draw[black, fill = black] (0.2, 0) circle (.02cm);
\draw[black, fill = black] (-0.2, 0) circle (.02cm);
\path[pyred, draw, line width = 1pt, postaction = decorate, decoration={markings,
			mark=at position 0.4 with {\arrow[line width=1pt]{>}}}] (-0.1, 0) arc (0:-360:0.1) -- (-0.1, 0);
\end{tikzpicture}}}\,,
\end{equation}
with the additional factor $i$ imposing the correct closed contour orientation. Without branch cuts in the energy domain, the additional identity $\G_{--}(k; t_1, t_2) = -\G_{++}(-k; t_1, t_2)$---which can be found directly from the frequency-space representation using $\frac{1}{(\omega^2 - k^2)_{i\epsilon}} \to \frac{1}{\omega^2-k^2-i\epsilon}$ as $k\to -k$ or from the explicit split representation in terms of mode functions---enables us to interpret the combination of propagators~(\ref{eq: propagator identity}) as cutting the internal line of a diagram on the same in-in branch since time integrals factorise. At the level of correlators, the additional input of unitarity relates individual diagrams to their complex conjugate version (with flipped external energies) \textit{on the same branch}. Consequently, the latest-time equation we derive should be understood as only a direct outcome of locality, as it is fundamentally determined by the structure of propagators. This general property remains valid in the presence of dissipation, fluctuations and non-unitary time evolution. 

\vskip 4pt
In de Sitter space, we have seen in Sec.~\ref{subsec: Spectral representation of massive cosmological propagators} that the effect of particle production amounts to double the usual poles while assigning distinct weights $e^{\pm\pi\mu}$ to the associated residues. From the spectral representation~(\ref{eq: massive propagator in de Sitter}), it is easy to observe that bulk-to-bulk propagators also satisfy the identity~(\ref{eq: propagator identity}), which, in terms of integration contours, is illustrated by
\begin{equation}
      \vcenter{\hbox{
\begin{tikzpicture}[line width=1. pt, scale=2]
\draw[black, -] (-0.5,0) -- (0.5,0) coordinate (xaxis);
\draw[black, -] (0,-0.4) -- (0,0.4) coordinate (yaxis);
\draw[black, fill = black] (0.2, 0.1) circle (.02cm);
\node at (0.35, 0.3) {$e^{-\pi\mu}$};
\draw[black, fill = black] (0.2, -0.1) circle (.02cm);
\node at (0.35, -0.3) {$e^{+\pi\mu}$};
\draw[black, fill = black] (-0.2, 0.1) circle (.02cm);
\node at (-0.35, 0.3) {$e^{+\pi\mu}$};
\draw[black, fill = black] (-0.2, -0.1) circle (.02cm);
\node at (-0.35, -0.3) {$e^{-\pi\mu}$};
\path[pyred, draw, line width = 1pt, postaction = decorate, decoration={markings,
			mark=at position 0.7 with {\arrow[line width=1pt]{>}}}] (-0.4, 0) -- (0.4, 0);
\end{tikzpicture}}}
\hspace*{0.22cm} - 
\vcenter{\hbox{
\begin{tikzpicture}[line width=1. pt, scale=2]
\draw[black, -] (-0.5,0) -- (0.5,0) coordinate (xaxis);
\draw[black, -] (0,-0.4) -- (0,0.4) coordinate (yaxis);
\draw[black, fill = black] (0.2, 0.1) circle (.02cm);
\node at (0.35, 0.3) {$e^{+\pi\mu}$};
\draw[black, fill = black] (0.2, -0.1) circle (.02cm);
\node at (0.35, -0.3) {$e^{-\pi\mu}$};
\draw[black, fill = black] (-0.2, 0.1) circle (.02cm);
\node at (-0.35, 0.3) {$e^{-\pi\mu}$};
\draw[black, fill = black] (-0.2, -0.1) circle (.02cm);
\node at (-0.35, -0.3) {$e^{+\pi\mu}$};
\path[pyred, draw, line width = 1pt, postaction = decorate, decoration={markings,
			mark=at position 0.7 with {\arrow[line width=1pt]{>}}}] (-0.4, 0) -- (0.4, 0);
\end{tikzpicture}}}
=
        \vcenter{\hbox{
\begin{tikzpicture}[line width=1. pt, scale=2]
\draw[black, -] (-0.5,0) -- (0.5,0) coordinate (xaxis);
\draw[black, -] (0,-0.4) -- (0,0.4) coordinate (yaxis);
\draw[black, fill = black] (0.2, 0) circle (.02cm);
\node at (0.35, 0.3) {$e^{+\pi\mu}$};
\draw[black, fill = black] (-0.2, 0) circle (.02cm);
\node at (-0.35, 0.3) {$e^{-\pi\mu}$};
\path[pyred, draw, line width = 1pt, postaction = decorate, decoration={markings,
			mark=at position 0.4 with {\arrow[line width=1pt]{>}}}] (0.3, 0) arc (0:-360:0.1) -- (0.3, 0);
\path[pyred, draw, line width = 1pt, postaction = decorate, decoration={markings,
			mark=at position 0.4 with {\arrow[line width=1pt]{>}}}] (-0.1, 0) arc (0:-360:0.1) -- (-0.1, 0);
\end{tikzpicture}}}
\hspace*{0.22cm} - 
\vcenter{\hbox{
\begin{tikzpicture}[line width=1. pt, scale=2]
\draw[black, -] (-0.5,0) -- (0.5,0) coordinate (xaxis);
\draw[black, -] (0,-0.4) -- (0,0.4) coordinate (yaxis);
\draw[black, fill = black] (0.2, 0) circle (.02cm);
\node at (0.35, 0.3) {$e^{-\pi\mu}$};
\draw[black, fill = black] (-0.2, 0) circle (.02cm);
\node at (-0.35, 0.3) {$e^{+\pi\mu}$};
\path[pyred, draw, line width = 1pt, postaction = decorate, decoration={markings,
			mark=at position 0.4 with {\arrow[line width=1pt]{>}}}] (0.3, 0) arc (0:-360:0.1) -- (0.3, 0);
\path[pyred, draw, line width = 1pt, postaction = decorate, decoration={markings,
			mark=at position 0.4 with {\arrow[line width=1pt]{>}}}] (-0.1, 0) arc (0:-360:0.1) -- (-0.1, 0);
\end{tikzpicture}}}\,.
\end{equation}
Crucially, note that particle production poles change their respective weight under complex conjugation. Sending $\mu\to\infty$, one immediately recovers the flat-space limit~(\ref{eq: flat-space propagator identity integration contour}).\footnote{The combination $\G_K \equiv \G_{++}+\G_{--}$ is nothing but the standard Keldysh propagator. The advanced, retarded and Keldysh propagators can be recovered after performing a rotation in the $\varphi_\pm$ field basis, introducing a new pair of fields $\varphi_r = \tfrac{1}{\sqrt{2}}(\varphi_+ + \varphi_-)$ and $\varphi_a = \tfrac{1}{\sqrt{2}}(\varphi_+ - \varphi_-)$, which is referred to the Keldysh basis~\cite{Keldysh:1964ud}. The advantage of this basis is that the physical interpretation of various propagators become transparent. Indeed, $\G_R$ and $\G_A$ characterise the dynamical properties of the particles in the system under consideration, while $\G_K$ characterises their statistical distribution and encodes the mean particle density~\cite{Kamenev_2011}.} The identity~(\ref{eq: propagator identity}) can of course be recovered from the form of propagators written with the time-ordered usual representation~(\ref{eq: SK propagators}) using $\Theta(t_1-t_2)+\Theta(t_2-t_1)=1$. Here, we have derived this identity from an off-shell perspective.

\subsection{Largest-time equation}

Our aim is to translate the propagator identity~(\ref{eq: propagator identity}) at the level of correlators, resulting in an identity satisfied by individual branch contributions of cosmological correlators. In the context of flat-space scattering amplitudes, this identity is known as the largest-time equation~\cite{VELTMAN1963186}.\footnote{Consider $F(\{x_i\})$ an amputated (real-space) Feynman diagram with $n$ vertices in flat space. Let us perform the following manipulations on this diagram:
\begin{itemize}
    \item Duplicate the Feynman diagram $2^n$ times by colouring vertices either in black or in white in all possible ways.
    \item For each vertex, a black-coloured vertex brings a factor $i$, and a white-coloured one brings a factor $(-i)$.
    \item The propagator between two black vertices is the usual Feynman propagator $\Delta_F(x-y) = \Theta(x^0-y^0)\Delta^+(x-y) + \Theta(y^0-x^0)\Delta^-(x-y)$, where $\Delta^\pm(x-y)$ are the usual positive and negative frequency commutation functions, while the propagator between two white vertices is taken to be $\Delta_F^*(x-y)$. The propagator between black and white vertices is replaced by $\Delta^+(x-y)$, and $\Delta^-(x-y)$ for white and black vertices. Clearly, the all-black diagram is the usual Feynman diagram while the all-white one is its complex conjugate. 
\end{itemize}
The flat-space largest-time equation states that the sum of all these contributions must vanish
\begin{equation}
    F(\{x_i\}) + F^*(\{x_i\}) + \bm{F}(\{x_i\}) = 0\,,
\end{equation}
where $F$ denotes the usual Feynman diagram, $F^*$ its complex conjugate, and $\bm{F}$ all other $2^n-2$ contributions. This is often taken as the starting point to derive recursion relations, see e.g.~\cite{Berends:1987me, Britto:2004ap, Bern:2011qt}, or cutting rules, see e.g.~\cite{Cutkosky:1960sp, Eden:1966dnq, VELTMAN1963186}. From the above colouring rules, it comes at no surprise that the Schwinger-Keldysh diagrammatic rules, and therefore cosmological correlators, are the most suited objects to phrase the largest-time equation.
}

\paragraph{Graph.} Computing correlators boils down to evaluating a set of multi-nested time integrals, whose number equals the number of vertices. Thus, instead of working with correlators, we define a \textit{graph} with $V$ vertices and $N$ internal lines by its collection of external energies $\{E_i\}$ ($i = 1, \ldots, V$) and internal energies $\{K_j\}$ ($j = 1, \ldots, N$). The resulting graph integrals are found after removing all overall factors of $1/2k$ from bulk-to-boundary and bulk-to-bulk propagators, hence defining the rescaled propagators
\begin{equation}
    \tilde{\K}_{\sf{a}}(k; t) \equiv 2k\, \K_{\sf{a}}(k; t)\,, \quad \tilde{\G}_{\sf{ab}}(k; t, t') \equiv 2k\, \G_{\sf{ab}}(k; t, t')\,,
\end{equation}
and overall coupling constants. We also multiply each graph integral by $(-i)^V$. Of course, in practice, the external energies are function of external momentum magnitudes. The main motivation is that the number of bulk-to-boundary propagators attached to a vertex, which we take to be massless in flat space and conformally coupled in de Sitter, does not matter, as they satisfy the trivial relation
\begin{equation}
    \prod_{i=1}^n \tilde{\K}_{\sf{a}}(k_i; t) = \tilde{\K}_{\sf{a}}\left(\sum_{i=1}^n k_i; t\right)\,,
\end{equation}
where $n$ is the number of external legs of a vertex. The sum of all external momentum magnitudes is defined to be the energy of the vertex, i.e.~$k_1 + \ldots + k_n = E$. Diagrammatically, we represent a graph the same way as a correlator (with each vertex carrying a Schwinger-Keldysh index) albeit with external legs removed, and instead labelling each vertex with its energy. For example, a one-site graph integral in flat space reads
\begin{equation}
    \raisebox{0pt}{
\begin{tikzpicture}[line width=1. pt, scale=2]
\draw (0, 0) circle (.05cm) node[above=1mm] {$E$};
\fill[black] (0,0) -- (90:0.05) arc (90:270:0.05) -- cycle;
\end{tikzpicture} 
} \equiv (-i)\, \sum_{{\sf{a}}=\pm} {\sf{a}}\int_{-\infty^{\sf{a}}}^0\d t \, e^{i {\sf{a}}Et}\,,
\end{equation}
which represents any contact correlator with arbitrary number of external legs. We also label internal lines with their corresponding internal energies.

\paragraph{A simple illustrative example.} Before stating the general largest-time equation, we illustrate the consequence of the identity~(\ref{eq: propagator identity}) on a simple example, namely the two-site chain. Even though the precise nature of the field $\varphi$, interactions, and the background is unimportant for what follows, we consider polynomial self-interactions in flat space and set coupling constants to unity. 

\vskip 4pt
First, let us consider the sum of both time-ordered contributions with flipped external energies on the negative branch
\begin{equation}
    \raisebox{0pt}{
\begin{tikzpicture}
[line width=1. pt, scale=2]
\draw[fill=black] (0, 0) circle (.05cm) node[above=0.5mm] {$E_1$};
\draw[fill=black] (1, 0) circle (.05cm) node[above=0.5mm] {$E_2$};
\draw[black] (0.05, 0) -- node[above] {$K$} (1, 0);
\end{tikzpicture} 
} + 
\raisebox{0pt}{
\begin{tikzpicture}
[line width=1. pt, scale=2]
\draw[fill=white] (0, 0) circle (.05cm) node[above=0.5mm] {$-E_1$};
\draw[fill=white] (1, 0) circle (.05cm) node[above=0.5mm] {$-E_2$};
\draw[black] (0.05, 0) -- node[above] {$K$} (0.95, 0);
\end{tikzpicture} 
}\,.
\end{equation}
It is given by 
\begin{equation}
    -\int_{-\infty^+}^0 \d t_1 \int_{-\infty^+}^0 \d t_2 \, e^{i E_1 t_1} \left[\tilde{\G}_{++}(K; t_1, t_2) + \tilde{\G}_{--}(K; t_1, t_2)\right] e^{i E_2 t_2}\,.
\end{equation}
Note that flipping external energies amounts to exchanging both in-in branches. As such, the $i\epsilon$ prescription that makes the integrals converge needs to be modified. Therefore, we have also deformed the lower bound of the time integrals accordingly so that both graph integrals can be combined. Using the propagator identity~(\ref{eq: propagator identity}) and flipping the external energies (and the corresponding lower bound of the time integral) for each integral appearing on the negative branch, we directly obtain
\begin{equation}
\label{eq: simple example factorised contributions}
    -\int_{-\infty^+}^0 \d t_1 \int_{-\infty^-}^0 \d t_2 \, e^{i E_1 t_1} \tilde{\G}_{+-}(K; t_1, t_2) e^{-i E_2 t_2} - \int_{-\infty^-}^0 \d t_1 \int_{-\infty^+}^0 \d t_2 \, e^{-i E_1 t_1} \tilde{\G}_{-+}(K; t_1, t_2) e^{i E_2 t_2}\,,
\end{equation}
which we recognise as 
\begin{equation}
    -\raisebox{0pt}{
\begin{tikzpicture}
[line width=1. pt, scale=2]
\draw[fill=black] (0, 0) circle (.05cm) node[above=0.5mm] {$E_1$};
\draw[fill=white] (1, 0) circle (.05cm) node[above=0.5mm] {$-E_2$};
\draw[black] (0.05, 0) -- node[above] {$K$} (0.95, 0);
\end{tikzpicture} 
} - 
\raisebox{0pt}{
\begin{tikzpicture}
[line width=1. pt, scale=2]
\draw[fill=white] (0, 0) circle (.05cm) node[above=0.5mm] {$-E_1$};
\draw[fill=black] (1, 0) circle (.05cm) node[above=0.5mm] {$E_2$};
\draw[black] (0.05, 0) -- node[above] {$K$} (0.95, 0);
\end{tikzpicture} 
}\,.
\end{equation}
Finally, let us group the factorised contributions in~(\ref{eq: simple example factorised contributions}) with the nested ones on the same side. The resulting diagrammatic formula reads
\begin{equation}
\raisebox{0pt}{
    \begin{tikzpicture}
[line width=1. pt, scale=2]
\draw[fill=black] (0, 0) circle (.05cm) node[above=0.5mm] {$E_1$};
\draw[fill=black] (1, 0) circle (.05cm) node[above=0.5mm] {$E_2$};
\draw[black] (0.05, 0) -- node[above] {$K$} (1, 0);
\end{tikzpicture} 
} + 
\raisebox{0pt}{
\begin{tikzpicture}
[line width=1. pt, scale=2]
\draw[fill=white] (0, 0) circle (.05cm) node[above=0.5mm] {$-E_1$};
\draw[fill=white] (1, 0) circle (.05cm) node[above=0.5mm] {$-E_2$};
\draw[black] (0.05, 0) -- node[above] {$K$} (0.95, 0);
\end{tikzpicture} 
}
    +\raisebox{0pt}{
\begin{tikzpicture}
[line width=1. pt, scale=2]
\draw[fill=black] (0, 0) circle (.05cm) node[above=0.5mm] {$E_1$};
\draw[fill=white] (1, 0) circle (.05cm) node[above=0.5mm] {$-E_2$};
\draw[black] (0.05, 0) -- node[above] {$K$} (0.95, 0);
\end{tikzpicture} 
} +
\raisebox{0pt}{
\begin{tikzpicture}
[line width=1. pt, scale=2]
\draw[fill=white] (0, 0) circle (.05cm) node[above=0.5mm] {$-E_1$};
\draw[fill=black] (1, 0) circle (.05cm) node[above=0.5mm] {$E_2$};
\draw[black] (0.05, 0) -- node[above] {$K$} (0.95, 0);
\end{tikzpicture} 
} = 0\,.
\end{equation}
Note that in this simple flat-space example, the rotation direction to negative energies does not matter as the mode functions are analytic. However, for massive fields in de Sitter space, one needs to specify the analytic continuation to negative energies so that it does not cross any branch cuts. In particular, energies should be rotated in opposite directions for the two branches
\begin{equation}
    \tilde{\K}_+(e^{-i\pi} k; t) = \tilde{\K}_-(k; t) \,, \quad \tilde{\K}_-(e^{+i\pi} k; t) = \tilde{\K}_+(k; t)\,.
\end{equation}
In most of the cases, we are interested in graphs with analytic external mode functions so that this prescription to evade the branch cut and define negative energies remains implicit.

\paragraph{Largest-time equation.} The previous simple example actually generalises to arbitrary graphs. In words, \textit{the sum of all individual graph contributions with negative external energies on the negative branch vanishes}. More formally and at tree-level, individual graph integral contributions can be written as
\begin{equation}
    \I_{{\sf{a}}_1 \ldots {\sf{a}}_V}\left(\{E_i\}, \{K_j\}\right) \equiv (-i)^V \int \prod_{i=1}^V \d t_i\, f_i(t_i) \, \tilde{\K}_{{\sf{a}}_i}(E_i; t_i) \prod_{j=1}^I \tilde{\G}_{{\sf{a}}_j {\sf{a}}_{j+1}}(K_j; t_j, t_{j+1})\,,
\end{equation}
where $V$ is the number of vertices, $I \equiv V-1$ is the number of internal lines (at tree-level), $E_i$ and $K_j$ denote external and internal energies, respectively, and $f_i(t_i)$ are form factors associated to each vertex that account for possible (time-dependent) couplings. We leave the dependence on momenta implicit. The labels ${\sf{a}}_j$ and ${\sf{a}}_{j+1}$ are to be understood as two consecutive Schwinger-Keldysh vertex indices that match those of the bulk-to-boundary propagators ${\sf{a}}_i$ (for ${\sf{a}}_j$) and ${\sf{a}}_{i+1}$ (for ${\sf{a}}_{j+1}$). Taking the values $\pm$, they indicate on which in-in branch the corresponding vertices are located. With these definitions, the latest-time equation can be simply stated as the following schematic identity
\begin{eBox}
\begin{equation}
\label{eq: latest-time equation}
    \sum_{{\sf{a}}_1, \ldots, {\sf{a}}_V = \pm} \I_{{\sf{a}}_1 \ldots {\sf{a}}_V} \left(\{{\sf{a}}_i E_i\}, \{K_j\}\right) = 0 \,.
\end{equation}
\end{eBox}
The above sum contains $2^V$ terms. In the presence of spatial derivative interactions, parity-violating interactions, or with external spinning fields, correlators can explicitly depend on external momenta $\bm{k}_i$, that can be contracted with e.g. polarisation tensors. In such cases, analytically continuing the energies $k_i \to -k_i$ also implies flipping the signs of the corresponding momenta $\bm{k}_i \to - \bm{k}_i$.\footnote{It should be pointed out that the latest-time equation can be understood as a consequence of the following combinatoric operator identity~\cite{Gillioz:2016jnn, Gillioz:2018kwh, Gillioz:2018mto, Meltzer:2020qbr}
\begin{equation}
    \sum_{k=0}^n (-1)^k \sum_{\sigma \in \Pi(k, n-k)} \bar{\text{T}} \left[\O(t_{\sigma_1}) \cdots \O(t_{\sigma_k})\right] \, \text{T} \left[\O(t_{\sigma_{k+1}}) \cdots \O(t_{\sigma_{n}})\right] = 0\,,
\end{equation}
where $\bar{\text{T}}$ and $\text{T}$ are the (anti-)time-ordering operators, and $\Pi(k, n-k)$ denotes the set of bi-partitions of $\{1, \ldots, n\}$ with size $k$ and $n-k$. This ``operator optical theorem" can be proved by induction. This means that the identity relating various in-in branches holds more generally at the operator level.}

\paragraph{Proof at tree-level.} To prove this identity for tree-level graphs, we use an inductive approach, as any tree graph can be constructed iteratively by successively adding internal lines. We follow the same lines as in the original work~\cite{VELTMAN1963186} where Cutkosky rules were derived for amputated Feynman diagrams, also see~\cite{Bourjaily:2020wvq, Meltzer:2020qbr, Baumann:2021fxj}.

\vskip 4pt
As seen above, the largest-time equation is satisfied for the simple two-site graph (the even more simple case of a one-site graph is trivial). We now assume that the latest-time equation~(\ref{eq: latest-time equation}) is valid for a tree-level graph with $V$ vertices and prove that it still remains valid when we add an additional vertex to the graph. Therefore, given a graph with $V$ vertices, we choose a vertex (say the farthest right one) and denote its corresponding external energy by $E_i$. Of course, this vertex carries $\pm$ labels. For what follows, it is useful to introduce some diagrammatic notation. We denote by the grey circle
\begin{equation}
    \vcenter{\hbox{
\begin{tikzpicture}
[line width=1. pt, scale=2]
\draw[fill=gray] (0, 0) circle (0.3cm);
\draw (0.3, 0) circle (.05cm) node[right=1mm] {$E_i$};
\fill[black] (0.3,0) -- (0.3, 0.05) arc (90:270:0.05) -- cycle;
\end{tikzpicture} 
    }}\,,
\end{equation}
the sum of all individual graph contributions with $V$ vertices, where the picked vertex $\raisebox{0pt}{
\begin{tikzpicture}[line width=1. pt, scale=2]
\draw (0, 0) circle (.05cm);
\fill[black] (0,0) -- (90:0.05) arc (90:270:0.05) -- cycle;
\end{tikzpicture} 
}$ has been brought forward. We now attach to it an additional vertex with external energy $E$ and we label the internal energy flowing through the bulk-to-bulk propagator by $K$. This increases the number of individual contributions by a factor 2, i.e.~from $2^V$ to $2^{V+1}$. Similarly to the simple illustrative example that we have treated above, we now consider the special combination of attaching both time-ordered bulk-to-bulk propagators with negative energies on the negative branch, which pictorially is represented by
\begin{equation}
    \vcenter{\hbox{
\begin{tikzpicture}
[line width=1. pt, scale=2]
\draw[fill=gray] (0, 0) circle (0.3cm);
\draw[fill=black] (0.3, 0) circle (.05cm) node[above right=0.1mm] {$E_i$};
\draw[fill=black] (1.3, 0) circle (.05cm) node[above=0.5mm] {$E$};
\draw[black] (0.35, 0) -- node[below] {$K$} (1.35, 0);
\end{tikzpicture} 
    }}
    +
    \vcenter{\hbox{
\begin{tikzpicture}
[line width=1. pt, scale=2]
\draw[fill=gray] (2.5, 0) circle (0.3cm);
\draw[fill=white] (2.8, 0) circle (.05cm) node[above right=0.1mm] {$-E_i$};
\draw[fill=white] (3.8, 0) circle (.05cm) node[above=0.5mm] {$-E$};
\draw[black] (2.85, 0) -- node[below] {$K$} (3.75, 0);
\end{tikzpicture} 
    }}\,.
\end{equation}
Note that we have added an additional factor of $i$ (which we include in the grey circle as it does not spoil the argument). Adding this vertex also brings an additional time integral. Flipping the external energies on the negative branch allows us to factor out all the information about external lines (and group both terms under same time integrals after also flipping the $i\epsilon$ prescription). Internal energies remain untouched. This manipulation isolates the bulk-to-bulk propagators to give
\begin{equation}
    \vcenter{\hbox{
\begin{tikzpicture}
[line width=1. pt, scale=2]
\draw[fill=gray] (0, 0) circle (0.3cm);
\end{tikzpicture} 
    }} \times \left(
    \vcenter{\hbox{
\begin{tikzpicture}
[line width=1. pt, scale=2]
\draw[fill=black] (0, 0) circle (.05cm);
\draw[fill=black] (1, 0) circle (.05cm);
\draw[black] (0, 0) -- (1, 0);
\end{tikzpicture} 
    }}
    +
    \vcenter{\hbox{
\begin{tikzpicture}
[line width=1. pt, scale=2]
\draw[fill=white] (0, 0) circle (.05cm);
\draw[fill=white] (1, 0) circle (.05cm);
\draw[black] (0.05, 0) -- (0.95, 0);
\end{tikzpicture} 
    }}
    \right)\,.
\end{equation}
This way, we can use the propagator identity~(\ref{eq: propagator identity}) to replace all time-ordered propagators with non-time-ordered ones, yielding
\begin{equation}
    \vcenter{\hbox{
\begin{tikzpicture}
[line width=1. pt, scale=2]
\draw[fill=gray] (0, 0) circle (0.3cm);
\end{tikzpicture} 
    }} \times \left(
    \vcenter{\hbox{
\begin{tikzpicture}
[line width=1. pt, scale=2]
\draw[fill=black] (0, 0) circle (.05cm);
\draw[fill=white] (1, 0) circle (.05cm);
\draw[black] (0.05, 0) -- (0.95, 0);
\end{tikzpicture} 
    }}
    +
    \vcenter{\hbox{
\begin{tikzpicture}
[line width=1. pt, scale=2]
\draw[fill=white] (0, 0) circle (.05cm);
\draw[fill=black] (1, 0) circle (.05cm);
\draw[black] (0.05, 0) -- (0.95, 0);
\end{tikzpicture} 
    }}
    \right)\,.
\end{equation}
These terms can then be redistributed to generate two individual graphs. At this stage, we recognise the two factorised contributions with flipped external energies on the negative branch, up to an overall minus sign coming from the Schwinger-Keldysh vertex indices. We obtain
\begin{equation}
    -\vcenter{\hbox{
\begin{tikzpicture}
[line width=1. pt, scale=2]
\draw[fill=gray] (0, 0) circle (0.3cm);
\draw[fill=black] (0.3, 0) circle (.05cm) node[above right=0.1mm] {$E_i$};
\draw[fill=white] (1.3, 0) circle (.05cm) node[above=0.5mm] {$-E$};
\draw[black] (0.35, 0) -- node[below] {$K$} (1.25, 0);
\end{tikzpicture} 
    }}
    -
    \vcenter{\hbox{
\begin{tikzpicture}
[line width=1. pt, scale=2]
\draw[fill=gray] (0, 0) circle (0.3cm);
\draw[fill=white] (0.3, 0) circle (.05cm) node[above right=0.1mm] {$-E_i$};
\draw[fill=black] (1.3, 0) circle (.05cm) node[above=0.5mm] {$E$};
\draw[black] (0.35, 0) -- node[below] {$K$} (1.35, 0);
\end{tikzpicture} 
    }}\,.
\end{equation}
Combining all terms on the same side, we finally obtain the desired diagrammatic identity. Since the simplest graph explicitly verifies this identity, the largest-time equation is proved for all tree-level graphs.

\paragraph{Loop-level diagrams.} Cuts, in the broad sense, are at the heart of modern flat-space scattering amplitude methods, especially at the loop level. For cosmological correlators, we will see here that the largest-time equation naturally generalises to loop graphs, as this identity stems from general properties of bulk-to-bulk propagators. 

\vskip 4pt
A loop diagram can essentially be generated from a tree-level diagram by merging two vertices. However, this process involves addressing two key points. First, the propagator identity~(\ref{eq: propagator identity}), which lies at the core of the largest-time equation at tree-level, must be extended to accommodate products of propagators. Second, loop diagrams contain unconstrained momenta flowing in the loops that need to be integrated over, which often results in divergences. As we are not primarily interested in loops, we will not explicitly perform loop integrals nor paying attention to regularising them. Instead, we will show how the largest-time equation holds at the loop level for a few simple graphs.

\vskip 4pt
\textit{One-site graph.} Let us consider the following simplest one-site graph at one-loop level with the vertex energy $E$
\begin{equation}
    \vcenter{\hbox{
\begin{tikzpicture}
[line width=1. pt, scale=2]
\draw[fill=white] (0, 0) circle (0.3cm);
\draw (0.3, 0) circle (.05cm) node[right=1mm] {$E$};
\fill[black] (0.3,0) -- (0.3, 0.05) arc (90:270:0.05) -- cycle;
\end{tikzpicture} 
    }}\,.
\end{equation}
This diagram consists of two distinct contributions that are complex conjugate to each other. As usual, we sum these contributions after flipping the energy on the negative branch, resulting in
\begin{equation}
    \vcenter{\hbox{
\begin{tikzpicture}
[line width=1. pt, scale=2]
\draw[fill=white] (0, 0) circle (0.3cm);
\draw[fill=black] (0.3, 0) circle (.05cm)node[right=1mm] {$E$};
\end{tikzpicture} 
    }}
    +
    \vcenter{\hbox{
\begin{tikzpicture}
[line width=1. pt, scale=2]
\draw[fill=white] (0, 0) circle (0.3cm);
\draw[fill=white] (0.3, 0) circle (.05cm)node[right=1mm] {$-E$};
\end{tikzpicture} 
    }}
    =
    -i\int_{-\infty^+}^0\d t \, e^{iE t} \int\frac{\d^3\bm{\ell}}{(2\pi)^3} \left[\tilde{\G}_{++}(\ell; t, t) - \tilde{\G}_{--}(\ell; t, t)\right]\,.
\end{equation}
This combination of graphs vanishes as both propagators are indistinguishable at equal times, i.e.~$\tilde{\G}_{++}(\ell; t, t) = \tilde{\G}_{--}(\ell; t, t)$. Notice that this straightforwardly extends to all one-site graphs to any loop order.

\vskip 4pt
\textit{Two-site graph.} We then consider the two-site graph at one-loop level
\begin{equation}
    \vcenter{\hbox{
\begin{tikzpicture}
[line width=1. pt, scale=2]
\draw[fill=white] (0, 0) circle (0.3cm);
\draw (-0.3, 0) circle (.05cm) node[left=1mm] {$E_1$};
\fill[black] (-0.3,0) -- (-0.3, 0.05) arc (90:270:0.05) -- cycle;
\draw (0.3, 0) circle (.05cm) node[right=1mm] {$E_2$};
\fill[black] (0.3,0) -- (0.3, 0.05) arc (90:270:0.05) -- cycle;
\end{tikzpicture} 
    }}\,.
\end{equation}
Each contribution contains a product of bulk-to-bulk propagators. The key insight is to realise that bulk-to-bulk propagators satisfy
\begin{equation}
\label{eq: generalised propagator identity}
    \prod_{i=1}^{L+1} \G_{++}(k_i, t_1, t_2) +\prod_{i=1}^{L+1} \G_{--}(k_i, t_1, t_2) = \prod_{i=1}^{L+1} \G_{+-}(k_i, t_1, t_2) +\prod_{i=1}^{L+1} \G_{-+}(k_i, t_1, t_2)\,,
\end{equation}
where $L$ denotes the number of loops. Of course, this relation is also valid for the rescaled (tilted) propagators. For $L=0$, we recover the identity~(\ref{eq: propagator identity}) that was used in the tree-level case. Proving this identity is simply a matter of manipulating and rearranging time orderings. For unequal times $t_1 \neq t_2$, one further requires the use of the trivial formula $\Theta(t_1-t_2)\Theta(t_2-t_1) = 0$, and at equal times $t_1 = t_2$, one should notice that the Wightman function is real $\G(k_i, t, t) = \G^*(k_i, t, t)$. The special value $\Theta(0)=1/2$ is used. At one-loop order $L=1$, we therefore have
\begin{equation}
    \begin{aligned}
    &\vcenter{\hbox{
\begin{tikzpicture}
[line width=1. pt, scale=2]
\draw[fill=white] (0, 0) circle (0.3cm);
\draw[fill=black] (-0.3, 0) circle (.05cm)node[left=1mm] {$E_1$};
\draw[fill=black] (0.3, 0) circle (.05cm)node[right=1mm] {$E_2$};
\end{tikzpicture} 
    }}
    +
    \vcenter{\hbox{
\begin{tikzpicture}
[line width=1. pt, scale=2]
\draw[fill=white] (0, 0) circle (0.3cm);
\draw[fill=white] (-0.3, 0) circle (.05cm)node[left=1mm] {$-E_1$};
\draw[fill=white] (0.3, 0) circle (.05cm)node[right=1mm] {$-E_2$};
\end{tikzpicture} 
    }} = -i \int_{-\infty^+}^0\d t_1\, e^{iE_1t_1} \d t_2 \,e^{iE_2t_2}\\
    &\times\int\frac{\d^3 \bm{\ell}}{(2\pi)^3}\left[\tilde{\G}_{++}(l; t_1, t_2) \tilde{\G}_{++}(|\bm{k}-\bm{l}|; t_1, t_2) + \tilde{\G}_{--}(l; t_1, t_2) \tilde{\G}_{--}(|\bm{k}-\bm{l}|; t_1, t_2)\right] \,.
    \end{aligned}
\end{equation}
Using~(\ref{eq: generalised propagator identity}), we recognise the sum of non-time-ordered contributions with flipped energies on the negative branch
\begin{equation}
    -\vcenter{\hbox{
\begin{tikzpicture}
[line width=1. pt, scale=2]
\draw[fill=white] (0, 0) circle (0.3cm);
\draw[fill=black] (-0.3, 0) circle (.05cm)node[left=1mm] {$E_1$};
\draw[fill=white] (0.3, 0) circle (.05cm)node[right=1mm] {$-E_2$};
\end{tikzpicture} 
    }}
    -
    \vcenter{\hbox{
\begin{tikzpicture}
[line width=1. pt, scale=2]
\draw[fill=white] (0, 0) circle (0.3cm);
\draw[fill=white] (-0.3, 0) circle (.05cm)node[left=1mm] {$-E_1$};
\draw[fill=black] (0.3, 0) circle (.05cm)node[right=1mm] {$E_2$};
\end{tikzpicture} 
    }}\,,
\end{equation}
albeit with an overall minus sign. Putting all terms on the same side, we recover the largest-time equation. Using~(\ref{eq: generalised propagator identity}) for $L>2$, this naturally generalises to any loop order.

\subsection{Application to simple diagrams}
\label{subsec: application to simple diagrams}

The primary challenge in performing perturbative calculations of cosmological correlators lies in the presence of nested time integrals, which are generally difficult to evaluate. However, as we have seen previously, time-ordering can be interpreted as a specific choice of contour to set the exchanged fields on shell. We can do this by considering a suitable integral representation for the bulk-to-bulk propagators, so that the off-shell propagators appear explicitly factorised.

\vskip 4pt
Here, before turning to correlators in a cosmological background which are of primary interest in Sec.~\ref{sec: Massive Exchange Correlator in de Sitter}, we first consider correlators in flat space.\footnote{In flat-space, we are interested in computing equal-time correlators at a fixed time
$t = 0$ so that interactions are building up during the time interval $-\infty < t < 0$. For concreteness, we consider a theory containing a single self-interacting massless scalar field $\varphi$. At tree level, the generalisation to (possible different) massive fields is straightforward after setting $k\to \sqrt{k^2+m^2}$. However, adding a mass to a mediated particle inside a loop is much more complicated.} In this case, the corresponding mode functions are analytic in the time (or energy) domain, and the frequency-space representation of the bulk-to-bulk propagator takes the exact same form as the usual Feynman propagator. Therefore, conventional flat-space methods can be used to compute correlators and reveal their analytic structure. In practice, once the factorised time integrals are evaluated, the remaining frequency integral can be finished with the residue theorem. We illustrate this off-shell factorisation with the simplest of all exchanged diagrams, and explicitly show that the computed correlators consistently satisfy the largest-time equation.

\begin{figure}[h!]
    \hspace*{0.5cm}
    \includegraphics[width=0.5\textwidth]{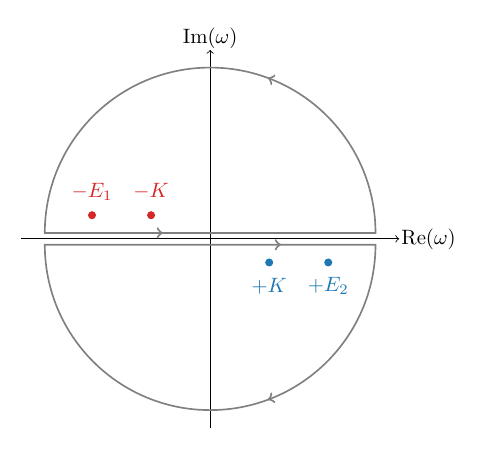}
   \caption{Illustration of the analytic structure of the integrand of $\I_{++}$ in the complex $\omega$ plane. Different choices of closing the contour, or equivalently different time orderings, relate \textcolor{pyblue}{physical poles} in the lower-half plane to \textcolor{pyred}{unphysical poles} in the upper-half plane.}
  \label{fig: flat-space 2-site chain analytic structure}
\end{figure}

\paragraph{Two-site chain.} Let us start with the simplest two-chain diagram\footnote{Concretely, for the interaction $\L = -\tfrac{g}{3!}\varphi^3$, the $s$-channel tree-level exchange correlator reads 
\begin{equation}
    \braket{\varphi_{\bm{k}_1} \varphi_{\bm{k}_2} \varphi_{\bm{k}_3} \varphi_{\bm{k}_4}}' = \frac{g^2}{16k_1k_2k_3k_4} \sum_{{\sf{a}, {\sf{b}}} = \pm} \I_{\sf{ab}}(E_1=k_{12}, E_2=k_{34}, K=s)\,,
\end{equation}
where $s = |\bm{k}_1+\bm{k}_2|$ is the magnitude of the exchange momentum, and $k_{ij} = k_i+k_j$.}
\begin{equation}
    \hspace*{-0.5cm}
    \raisebox{0pt}{
\begin{tikzpicture}
[line width=1. pt, scale=2]
\draw (0, 0) circle (.05cm) node[above=1mm] {$E_1$};
\fill[black] (0,0) -- (0, 0.05) arc (90:270:0.05) -- cycle;
\draw (1, 0) circle (.05cm) node[above=1mm] {$E_2$};
\fill[black] (1,0) -- (1, 0.05) arc (90:270:0.05) -- cycle;
\draw[black] (0.05, 0) -- node[above] {$K$} (1, 0);
\end{tikzpicture} 
} = \sum_{{\sf{a}, {\sf{b}}} = \pm} \I_{\sf{ab}} \,,\,\, \I_{\sf{ab}} = -{\sf{ab}} \int_{-\infty^{\sf{a}}}^0 \d t_1 \int_{-\infty^{\sf{b}}}^0 \d t_2 \, e^{i{\sf{a}}E_1 t_1}\tilde{\G}_{\sf{ab}}(K; t_1, t_2) e^{i{\sf{b}}E_2 t_2}\,.
\end{equation}
In order to compute the time integrals, the $i\epsilon$ prescription in the in-in contour can be translated to a shift in energy through the formula
\begin{equation}
\label{eq: flat-space time integral pole prescription}
    \int_{-\infty^\pm}^0 \d t \, e^{\pm izt} = \int_{-\infty}^0 \d t \, e^{\pm i(z \mp i\epsilon)t} = \frac{\mp i}{z\mp i\epsilon}\,,
\end{equation}
for a real positive energy $z>0$. This is equivalent to Wick rotating time, as is usually done. However, it gives the correct pole prescription for the later frequency integration when computing the nested contribution. In the limit $\epsilon\to0$, the unnested integral is simply given by
\begin{equation}
    \I_{+-} = \int_{-\infty^+}^0 \d t_1 e^{i(E_1+K)t_1} \int_{-\infty^-}^0 \d t_2 e^{-i(E_2+K)t_2} = \frac{1}{E_L E_R}\,,
\end{equation}
where $E_L \equiv E_1+K$ and $E_R \equiv E_2+K$ are the sum of energies at the left and right vertices. Note that this contribution has only partial energy poles when $E_{L, R}\to0$, which emerge from the factorised form. Since the result is purely real, we have $\I_{-+} = \I_{+-}$. Similarly, introducing the frequency-space representation of the time-ordered bulk-to-bulk propagator~(\ref{eq: flat-space bulk-to-bulk propagator}) and performing the time integrals using~(\ref{eq: flat-space time integral pole prescription}), the time-ordered contribution reads
\begin{equation}
    \I_{++} = -2K\int_{-\infty}^{+\infty}\frac{\d\omega}{2i\pi} \frac{1}{(E_1+\omega-i\epsilon)(\omega^2-K^2)_{i\epsilon}(E_2-\omega-i\epsilon)}\,.
\end{equation}
This representation makes it explicit that both time integrals factorise and that time ordering is encoded in the additional frequency integral. The symmetry under $E_1 \leftrightarrow E_2$ can be restored after changing variables $\omega \to -\omega$. The analytic structure of the integrand in the complex $\omega$ plane exhibits several interesting features that we show in Fig.~\ref{fig: flat-space 2-site chain analytic structure}: (i) the pole $\omega = +E_2-i\epsilon$ is the off-shell collinear (folded) limit, (ii) the pole $\omega = -E_1+i\epsilon$ is the off-shell (left) partial energy pole, and (iii) the poles $\omega = \pm (K-i\epsilon)$ are the on-shell conditions. Notice that poles that can only be reached after analytically continuing the energies to the unphysical kinematic configuration (after setting the exchanged particle on shell) are located in the upper-half complex $\omega$ plane. Therefore, selecting for example the \textit{physical} on-shell condition requires closing the contour in the lower-half complex plane, picking the residues at $\omega=+K-i\epsilon$ and $\omega=+E_2-i\epsilon$, we obtain
\begin{equation}
    \I_{++} = \frac{1}{E}\left(\frac{1}{E_L} + \frac{1}{E_R}\right)\,,
\end{equation}
where $E \equiv E_1+E_2$ is the total energy of the graph. Of course, closing the contour upwards yields the same result. This shows that partial energy singularities reached for \textit{unphysical} kinematic configurations are related to folded singularities reached for \textit{physical} ones by different time ordering choices. In this representation, the total energy pole $E\to0$ is made explicit through the identity
\begin{equation}
\label{eq: apart identity}
    \frac{1}{(E_1+\omega-i\epsilon)(E_2-\omega-i\epsilon)} = \frac{1}{E}\left(\frac{1}{E_1+\omega-i\epsilon} + \frac{1}{E_2-\omega-i\epsilon}\right)\,,
\end{equation}
and partial energy poles are reconstructed from the frequency integral. By reality of the result, we have $\I_{--} = \I_{++}$. Summing all contributions, the final result reads
\begin{equation}
\label{eq: flat-space tree-level 4pt function}
    \I(\{E_i\}, K) = \frac{f(\{E_i\}, K)}{E E_L E_R}\,, \quad \text{with} \quad f(\{E_i\}, K) = 4(E_1+E_2+K)\,.
\end{equation}
This graph integral is a rational function of the energies and has a singularity when the total energy vanishes $E\to0$, and when partial energies flowing into the left or right vertices add up to zero $E_{L, R}\to0$, see e.g.~\cite{Arkani-Hamed:2017fdk, Arkani-Hamed:2018bjr, Benincasa:2018ssx, Arkani-Hamed:2018kmz, Baumann:2021fxj} for more details. The residue at the total energy poles is the corresponding scattering amplitude~\cite{Maldacena:2011nz, Raju:2012zr}. Indeed, using the identity $E_L+E_R=2K$ in the limit $E\to0$, the graph integral can be written
\begin{equation}
    \lim\limits_{E \to 0} \, \tfrac{1}{4K}\,\I(\{E_i\}, K) = \frac{\A}{E}\,,
\end{equation}
where $\A = \tfrac{1}{-\S}$ is the corresponding $s$-channel scattering amplitude with $\S=-(k_1^\mu + k_2^\mu)^2=E_1^2-K^2$ being the usual Mandelstam variable. Note that in order to extract the scattering amplitude in Eq.~(\ref{eq: flat-space tree-level 4pt function}), we have imposed the energy conservation condition $E_1+E_2=0$---valid in the asymptotic past---to write $K^2-E_1^2 = E_L E_R$. Eventually, the identity~(\ref{eq: apart identity}) makes it explicit that the graph integrals $\I_{\sf{ab}}$ satisfy the largest-time equation
\begin{equation}
    \I_{++}(E_1, E_2, K) + \I_{--}(-E_1, -E_2, K) + \I_{+-}(E_1, -E_2, K) + \I_{-+}(-E_1, E_2, K) = 0\,.
\end{equation}

\paragraph{Three-site chain.} We now examine the three-site chain, which presents additional complexities compared to the two-site chain. Notably, this requires evaluating two frequency integrals for the fully nested contribution. The corresponding graph reads
\begin{equation}
    \begin{aligned}
    \hspace*{-0.5cm}
    \raisebox{0pt}{
\begin{tikzpicture}
[line width=1. pt, scale=2]
\draw (0, 0) circle (.05cm) node[above=1mm] {$E_1$};
\fill[black] (0,0) -- (0, 0.05) arc (90:270:0.05) -- cycle;
\draw (1, 0) circle (.05cm) node[above=1mm] {$E_2$};
\fill[black] (1,0) -- (1, 0.05) arc (90:270:0.05) -- cycle;
\draw[black] (0.05, 0) -- node[above] {$K_1$} (1, 0);
\draw (2, 0) circle (.05cm) node[above=1mm] {$E_3$};
\fill[black] (2,0) -- (2, 0.05) arc (90:270:0.05) -- cycle;
\draw[black] (1.05, 0) -- node[above] {$K_2$} (2, 0);
\end{tikzpicture} 
} &= \sum_{{\sf{a}, {\sf{b}}, {\sf{c}}} = \pm} \I_{\sf{abc}} \,, \quad \text{with} \\
\I_{\sf{abc}} = i {\sf{abc}} \int_{-\infty^{\sf{a}}}^0 \d t_1 \int_{-\infty^{\sf{b}}}^0 \d t_2 \int_{-\infty^{\sf{c}}}^0 \d t_3 \, e^{i{\sf{a}}E_1t_1}&\tilde{\G}_{\sf{ab}}(K_1; t_1, t_2) e^{i{\sf{b}}E_2t_2} \tilde{\G}_{\sf{bc}}(K_2; t_2, t_3) e^{i{\sf{c}}E_3t_3}\,.
    \end{aligned}
\end{equation}
We have in total $2^3=8$ integrals to evaluate, among which half of them are related by complex conjugation. Using Eq.~(\ref{eq: flat-space time integral pole prescription}) and the previous result for evaluating a single frequency integral, the fully factorised and partially nested contributions read
\begin{equation}
    \begin{aligned}
        \I_{+-+} = \frac{-1}{E_L E_M E_R}\,, \quad \I_{++-} = \frac{-1}{\bar{E}_LE_R}\left(\frac{1}{E_L} + \frac{1}{E_M}\right)\,, \quad \I_{-++} = \frac{-1}{E_L \bar{E}_R} \left(\frac{1}{E_M} + \frac{1}{E_R}\right)\,,
    \end{aligned}
\end{equation}
where $E_L \equiv E_1+K_1, E_M \equiv E_2+K_1+K_2$ and $E_R \equiv E_3+K_2$ are the sum of energies at the left, middle and right vertices, respectively, and $\bar{E}_L \equiv E_1+E_2+K_2$ and $\bar{E}_R \equiv E_2+E_3+K_1$ are the energies of the joined two left and right vertices, respectively. Similar integrals were computed in~\cite{Arkani-Hamed:2017fdk} for the corresponding wavefunction coefficient. After using the frequency-space representation for the time-ordered bulk-to-bulk propagators and evaluating the time integrals, the fully nested contribution reads
\begin{equation}
    \begin{aligned}
        \I_{+++} = -(2K_1)(2K_2) &\int_{-\infty}^{+\infty} \frac{\d\omega_1}{2i\pi}\frac{\d\omega_2}{2i\pi} \frac{1}{(\omega_1^2-K_1^2)_{i\epsilon}(\omega_2^2-K_2^2)_{i\epsilon}} \\
        &\times \frac{1}{(E_1+\omega_1-i\epsilon)(E_2-\omega_1+\omega_2-i\epsilon)(E_3-\omega_2-i\epsilon)}\,.
    \end{aligned}
\end{equation}
This double integral can be evaluated after expanding the integrand into partial fractions, rendering the off-shell poles manifest, and then picking up residues. We obtain
\begin{equation}
    \begin{aligned}
        \I_{+++} = &-\frac{1}{E}\left(\frac{1}{E_2+E_3+K_1}-\frac{1}{E_2+E_3-K_1}\right)\left(\frac{1}{E_3+K_2}-\frac{1}{E_3-K_2}\right) \\
        &-\frac{1}{(E_3-K_2)(E_1+E_2+K_2)}\left(\frac{1}{E_2+K_1+K_2} -\frac{1}{E_2-K_1+K_2}\right) \\
        &- \frac{1}{(E_1+K_1)(E_2+E_3-K_1)}\left(\frac{1}{E_3+K_2}-\frac{1}{E_3-K_2}\right) \\
        &- \frac{1}{(E_1+K_1)(E_2-K_1+K_2)(E_3-K_2)}\,.
    \end{aligned}
\end{equation}
Notice that it is important to keep track of the correct $i\epsilon$ pole prescription to collect the correct residues. Eventually, since all contributions are real, the remaining ones are identical. Summing all contributions, the graph integral is
\begin{equation}
    \begin{aligned}
        &\I(\{E_i\}, \{K_i\}) = \frac{f(\{E_i\}, \{K_i\})}{E E_L \bar{E}_L E_M \bar{E}_R E_R}\,, \\
        \text{with} \quad f(\{E_i\}, \{K_i\}) &= -8\left[\bar{E}_LE_M \bar{E}_R + E_M(E_1\bar{E}_L+E_3 \bar{E}_R)+E_1E_3(\bar{E}_L+\bar{E}_R)\right]\,.
    \end{aligned}
\end{equation}
The fairly non-trivial form factor $f$ is well symmetric under $(E_1 \leftrightarrow E_3, K_1 \leftrightarrow K_2)$, and the final result displays the correct $E_L, E_M, E_R, \bar{E}_L, \bar{E}_R$ and $E$ poles, as expected. One can explicitly verify that individual contributions satisfy the largest-time equation
\begin{equation}
    \begin{aligned}
        0 &= \I_{+++}(E_1, E_2, E_3, K_1, K_2) + \I_{---}(-E_1, -E_2, -E_3, K_1, K_2) \\ &+ \I_{++-}(E_1, E_2, -E_3, K_1, K_2) + \I_{--+}(-E_1, -E_2, E_3, K_1, K_2) \\
        &+ \I_{-++}(-E_1, E_2, E_3, K_1, K_2) + \I_{+--}(E_1, -E_2, -E_3, K_1, K_2) \\
        &+ \I_{+-+}(E_1, -E_2, E_3, K_1, K_2) + \I_{-+-}(-E_1, E_2, -E_3, K_1, K_2)\,,
    \end{aligned}
\end{equation}
providing a non-trivial check of the final result. In principle, this procedure can be generalised to arbitrary tree-level diagrams. However, the number of required integrals and the complexity of multi-dimensional frequency integrals rapidly become intractable.

\paragraph{Conformal two-site chain in de Sitter.} As a last example, we consider the two-site chain of conformally coupled scalar fields (with mass $m^2=2H^2$) in de Sitter, whose computation conceptually follows the same line as in flat space, as the corresponding time-ordered bulk-to-bulk propagators can be recasted as frequency-space integrals.\footnote{It is possible to map correlators of conformally coupled fields to correlators of massive scalars with half-integer values of $\nu$ by applying a set of weight-shifting operators. This procedure follows from the property of the Hankel function that drastically simplifies for $\nu=n/2$ with $n$ integer ($n=1$ is the conformally coupled case).} The corresponding graph integrals are
\begin{equation}
    \hspace*{-0.5cm}
    \raisebox{0pt}{
\begin{tikzpicture}
[line width=1. pt, scale=2]
\draw (0, 0) circle (.05cm) node[above=1mm] {$E_1$};
\fill[black] (0,0) -- (0, 0.05) arc (90:270:0.05) -- cycle;
\draw (1, 0) circle (.05cm) node[above=1mm] {$E_2$};
\fill[black] (1,0) -- (1, 0.05) arc (90:270:0.05) -- cycle;
\draw[black] (0.05, 0) -- node[above] {$K$} (1, 0);
\end{tikzpicture} 
} = \sum_{{\sf{a}, {\sf{b}}} = \pm} \I_{\sf{ab}} \,,\,\, \I_{\sf{ab}} = -{\sf{ab}} \int_{-\infty^{\sf{a}}}^{\tau_0} \frac{\d\tau_1}{\tau_1^2} \int_{-\infty^{\sf{b}}}^{\tau_0} \frac{\d\tau_2}{\tau_2^2} \, e^{i{\sf{a}}E_1 \tau_1}\tilde{\G}_{\sf{ab}}(K; \tau_1, \tau_2) e^{i{\sf{b}}E_2 \tau_2}\,,
\end{equation}
where $\tau_0<0$ is a small late-time cutoff. Shifting the energy and neglecting analytic terms in the late-time limit, we use the formula
\begin{equation}
    \int_{-\infty^\pm}^{\tau_0} \frac{\d\tau}{\tau} e^{\pm i z \tau} = \int_{-\infty}^{\tau_0} \frac{\d\tau}{\tau} e^{\pm i (z\mp i \epsilon) \tau} = \log\left[\pm i (z\mp i\epsilon)\tau_0\right]\,,
\end{equation}
for a real positive energy $z>0$, to write the factorised contribution as
\begin{equation}
    \I_{+-} = \log(+i E_L \tau_0)\log(-iE_R\tau_0)\,.
\end{equation}
Similarly, after performing the time integrals for the nested contribution, we obtain
\begin{equation}
    \I_{++} = 2K \int_{-\infty}^{+\infty} \frac{\d\omega}{2i\pi} \frac{\log[i(E_1+\omega-i\epsilon)]\log[i(E_2-\omega-i\epsilon)]}{(\omega^2-K^2)_{i\epsilon}}\,.
\end{equation}
This integral, besides being formally divergent, is hard to evaluate because of the branch cuts spanning over $\omega \in (-E_1+i\epsilon-i\infty, -E_1+i\epsilon)$ and $\omega \in (E_2-i\epsilon+i\infty, E_2-i\epsilon)$ (choosing the log branch cut to be the principal one). These branch cuts notably cross the real axis. As we will see in Sec.~\ref{sec: Massive Exchange Correlator in de Sitter}, the presence of branch cuts, rather than poles, indicates particle production. In this scenario, the integrand can be transformed into a meromorphic function by transitioning to the complex mass plane instead of the complex frequency plane. However, for the special case of conformally coupled fields, as first noticed in~\cite{Arkani-Hamed:2015bza}, this integral can be written as an energy integral over the flat-space result
\begin{equation}
    \begin{aligned}
        \I_{++} &= 2K \int_{-\infty}^{+\infty}\frac{\d\omega}{2i\pi} \frac{1}{(\omega^2-K^2)_{i\epsilon}} \int_{E_1+\omega-i\epsilon}^\infty \int_{E_2-\omega-i\epsilon}^\infty \frac{\d x \d y}{xy} \\
        &= 2K \int_{E_1}^\infty\d x \int_{E_2}^\infty\d y \int_{-\infty}^{+\infty}\frac{\d\omega}{2i\pi} \frac{1}{(x+\omega-i\epsilon)(\omega^2-K^2)_{i\epsilon}(y-\omega-i\epsilon)}\,,
    \end{aligned}
\end{equation}
where we have shifted the kinematic integrals so that their limits do not depend on the off-shell frequency. We recognise the flat-space frequency integral that we previously computed. The energy integrals were first computed in~\cite{Arkani-Hamed:2015bza} and the result reads\footnote{In order to extract the divergence, one can subtract and add the term
\begin{equation}
    \int_{-\infty^+}^{\tau_0}\frac{\d\tau_1}{\tau_1} \frac{\d\tau_2}{\tau_2} e^{i(E_L\tau_1+E_R\tau_2)} = \log(iE_L \tau_0)\log(i E_R \tau_0) = \int_{E_1}^\infty\d x \int_{E_2}^\infty\d y \, \frac{1}{(x+K-i\epsilon)(y+K-i\epsilon)}\,,
\end{equation}
which can either be written as a product of logarithms (hence isolating the divergence) or as energy integrals that can be combined with the original ones. This effectively mimics the corresponding wavefunction coefficient calculation. The remaining convergent integral to be computed is
\begin{equation}
    \int_0^\infty \frac{\d x \d y}{(x+y+E)(x+E_L)(y+E_R)}\,.
\end{equation}
Sophisticated techniques to evaluate such integrals are detailed in~\cite{Hillman:2019wgh, De:2023xue, Arkani-Hamed:2023kig, Fan:2024iek}.}
\begin{equation}
    \begin{aligned}
        \I_{++} &= \Li_2\left(\frac{E-E_L}{E}\right) + \Li_2\left(\frac{E-E_R}{E}\right) + \log\left(\frac{E_L}{E}\right)\log\left(\frac{E_R}{E}\right) - \frac{\pi^2}{6} \\
        &- \log(i E_L \tau_0)\log(i E_R\tau_0)\,,
    \end{aligned}
\end{equation}
where $\Li_2$ is the dilogarithm function. When summing all contributions, the kinematic parts of divergent pieces cancel against each other thanks to the property $\log(iz)-\log(-iz)=-i\pi$ for $z<0$. Using the Euler's identity
\begin{equation}
    \Li_2(z)+\Li_2(1-z)+\log(z)\log(1-z)=\frac{\pi^2}{6}\,,
\end{equation}
both nested contributions combine to give
\begin{equation}
    \begin{aligned}
        \I_{++}(E_1, E_2, K) + \I_{--}(-E_1, -E_2, K) = &-\log[i(E_1-K)\tau_0]\log[i(E_2+K)\tau_0]\\
        &-\log[i(E_1+K)\tau_0]\log[i(E_2-K)\tau_0]\,,
    \end{aligned}
\end{equation}
which is exactly minus the combination $\I_{+-}(E_1, -E_2, K)+\I_{-+}(-E_1, E_2, K)$. Consequently, the graph integrals $\I_{\sf{ab}}$ satisfy the largest-time equation.

\section{Massive Exchange Correlator in de Sitter}
\label{sec: Massive Exchange Correlator in de Sitter}

We now turn to the case of the two-site chain of conformally coupled scalars mediated by the tree-level exchange of massive scalars in de Sitter. By explicitly performing the spectral integral, we derive a new closed-form expression for this correlator. As we will show, the branch cut in the energy domain due to particle production translates into a tower of poles in the complex mass domain. Summing over these residues leads to a new simple and partially resummed series solution.

\vskip 4pt
We are interested in computing the de Sitter tree-level exchange graph\footnote{Concretely, considering a conformally coupled scalar field $\varphi$ coupled to a massive scalar field $\phi$ by $\L_{\text{int}}/a^3(t) = -\tfrac{g}{2}\phi \varphi^2$ in de Sitter, the full correlator reads
\begin{equation}
    \braket{\varphi_{\bm{k}_1} \varphi_{\bm{k}_2} \varphi_{\bm{k}_3} \varphi_{\bm{k}_4}}' = \frac{g^2 \tau_0^4}{16k_1k_2k_3k_4} \, F(E_1 = k_{12}, E_2 = k_{34}, K = s) + t + u \text{ channels}\,,
\end{equation}
where we have introduced a late-time cutoff $\tau_0$ and defined $F \equiv \sum_{{\sf{a, b}}=\pm} F_{\sf{ab}}$. Notice that we have defined the integrals $F_{\sf{ab}}$ without the factor $2$ to ease comparison with the literature.}
\begin{equation}
\label{eq: 4pt function F definition}
    \begin{aligned}
    \raisebox{0pt}{
\begin{tikzpicture}
[line width=1. pt, scale=2]
\draw (0, 0) circle (.05cm) node[above=1mm] {$E_1$};
\fill[black] (0,0) -- (0, 0.05) arc (90:270:0.05) -- cycle;
\draw (1, 0) circle (.05cm) node[above=1mm] {$E_2$};
\fill[black] (1,0) -- (1, 0.05) arc (90:270:0.05) -- cycle;
\draw[black] (0.05, 0) -- node[above] {$K$} (1, 0);
\end{tikzpicture} 
} &= \sum_{{\sf{a}, {\sf{b}}} = \pm} F_{\sf{ab}} \,,\\
\text{with} \quad F_{\sf{ab}} = -{\sf{ab}}K \int_{-\infty^{\sf{a}}}^0 \frac{\d\tau'}{(-\tau')^{1/2}}e^{i{\sf{a}}E_1\tau'}&\int_{-\infty^{\sf{b}}}^0 \frac{\d\tau''}{(-\tau'')^{1/2}}e^{i{\sf{b}}E_2\tau''} \G_{\sf{ab}}(K; \tau', \tau'')\,.
    \end{aligned}
\end{equation}
Recall that $\G_{\sf{ab}}$ are the propagators for the canonically normalised field $\sigma(\tau, \bm{x}) \equiv (-\tau)^{-3/2} \phi(\tau, \bm{x})$, hence the unusual power for the conformal time within the integral. We define the usual dimensionless kinematic variables 
\begin{equation}
    u \equiv \frac{K}{E_1}\,, \quad v \equiv \frac{K}{E_2}\,,
\end{equation}
which lie in the unit disk, i.e.~$|u|, |v| \leq 1$, and the three-point function
\begin{equation}
\label{eq: 3pt function definition}
    F^{(3)}(z, \mu) = \tfrac{1}{\sqrt{2}} |\Gamma(\tfrac{1}{2}+i\mu)|^2 \, P_{i\mu-1/2}(z)\,,
\end{equation}
where $P_{i\mu-1/2}$ is the Legendre function defined in Eq.~(\ref{eq: Legendre P Q definitions}). The non-time-ordered contribution, for which both time integrals factorise, is therefore given by
\begin{equation}
\label{eq: F_+- contribution}
    F_{+-} = F^{(3)}(u^{-1}, \mu) \, F^{(3)}(v^{-1}, \mu)\,.
\end{equation}
We have used the useful formula~(\ref{eq: integral of Hankel}) to perform the time integrals. Notice that the Legendre $P$ function is real, i.e.~$[P_{i\mu-1/2}(z)]^* = P_{i\mu-1/2}(z)$ for $z\geq 0$ and $\mu$ real. Therefore, the on-shell three-point function $F^{(3)}(z, \mu)$ is real. As a consequence, we obtain the same expression for $F_{-+}$. However, anticipating what follows, the off-shell function $F^{(3)}(z, \nu)$ where the mass parameter $\nu$ is analytically continued to the complex plane is not real.

\subsection{Off-shell three-point function}
\label{subsec: Off-shell three-point function}

For pedagogical reasons, before computing the spectral integral for the time-ordered contribution $F_{++}$, we first study the following simpler integral
\begin{equation}
\label{eq: off-shell 3pt function}
    \bar{F}^{(3)}(z) = \int_{-\infty}^{+\infty} \d\nu \,\N_\nu \, \frac{F^{(3)}(z, \nu)}{(\nu^2-\mu^2)_{i\epsilon}}\,,
\end{equation}
with $z\geq 1$, and $F^{(3)}$ defined in Eq.~(\ref{eq: 3pt function definition}). The object $\bar{F}^{(3)}$ is typically proportional to $F_{++}$ in the folded limit $E_2\to K$ (or $E_1\to K$ by symmetry), equivalently $v\to 1$ (or $u\to 1$), and already exhibits an interesting analytic structure. To be able to perform the spectral integral, it is essential to understand the analytic structure of the function $F^{(3)}(z, \nu)$ in the complex $\nu$ plane, as well as its behaviour at infinity, to appropriately determine how to close the contour.

\begin{figure}[h!]
   \hspace*{0.5cm}
    \includegraphics[width=0.5\textwidth]{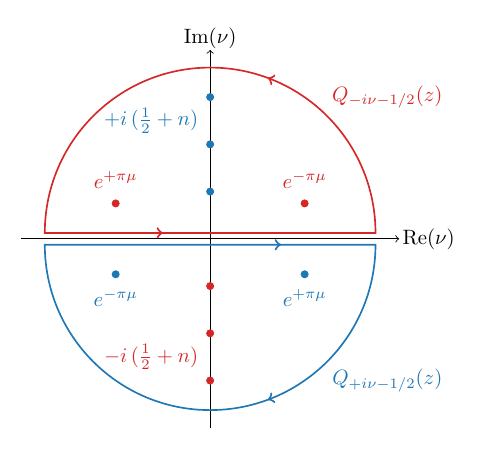}
   \caption{Illustration of the analytic structure of the off-shell three-point function integrand in~(\ref{eq: off-shell 3pt function}) in the complex $\nu$ plane, written in the Legendre $Q$ basis. The contour prescription avoids the infinite tower of poles lying on the imaginary axis and only picks up particle production poles.}
  \label{fig: off-shell 3pt function}
\end{figure}

\vskip 4pt
The crucial observation is that the off-shell three-point function $F^{(3)}$ is a meromorphic function in the complex $\nu$ plane. At a fixed kinematic configuration $z$, the function $F^{(3)}(z, \nu)$ has poles at $\nu = \pm i\, (\tfrac{1}{2}+n)$ with integer $n\geq0$, which corresponds to the poles of $|\Gamma(\tfrac{1}{2}+i\nu)|^2 = \Gamma(\tfrac{1}{2}+i\nu) \Gamma(\tfrac{1}{2}-i\nu)$, since the Legendre function $P$ is analytic in the entire complex $\nu$ plane. However, the Legendre $P$ representation of $F^{(3)}$ makes the large-$\nu$ asymptotic form not neat, i.e.~it depends on the phase of $z$. In order to disentangle both behaviours at infinity, we project the off-shell three point function onto the Legendre $Q$ basis using the connection formula~(\ref{eq: Legendre P/Q connection formula}), yielding
\begin{equation}
    F^{(3)}(z, \nu) = \frac{e^{-\tfrac{i\pi}{2}}}{\sqrt{2}\sinh(\pi\nu)} \left[Q_{-i\nu-1/2}(z) - Q_{+i\nu-1/2}(z)\right]\,.
\end{equation}
The function $Q_{\pm i\nu - 1/2}(z)$ has purely imaginary poles located at $\nu = \pm i\, (\tfrac{1}{2}+n)$ (which come from $\Gamma(\tfrac{1}{2}\pm i\mu)$ in Eq.~(\ref{eq: Legendre P Q definitions})), and should be closed in the lower (upper) complex plane since $Q_{\pm i\nu -1/2}(z) \sim e^{\mp i\nu\xi}$ with $\xi >0$. Therefore, when performing the spectral integral, we never need to pick up these poles because the contour is closed to avoid the infinite tower.\footnote{The factor $1/\sinh(\pi\nu) = |\Gamma(1+i\nu)|^2/(\pi\nu)$ also brings additional poles to the integrand. However, they are precisely cancelled by the density of states $\N_\nu$.} We illustrate the analytic structure of the integrand~(\ref{eq: off-shell 3pt function}) and the contour prescription in Fig~\ref{fig: off-shell 3pt function}. Eventually, picking up only the particle production poles, we obtain
\begin{equation}
\label{eq: integrated off-shell 3pt function}
    \bar{F}^{(3)}(z) = F^{(3)}(z e^{+i\pi}, \mu)\,,
\end{equation}
where we have used the analytic continuation of the Legendre function~(\ref{eq: Legendre P analytic continuation}) to re-express the result in the Legendre $P$ basis. As already seen in Sec.~\ref{subsec: Spectral representation of massive cosmological propagators}, the particle production pole structure projects ingoing modes on outgoing ones and vice versa. At the level of the three-point function, this amounts to rotate the kinematic configuration $z \to ze^{+i\pi}$. Actually, we show in App.~\ref{subsec: app dispersive integral} that the function $F^{(3)}(z, \mu)$ and its analytic continuation $F^{(3)}(ze^{+i\pi}, \mu)$, viewed as functions of $z$ for fixed $\mu$, are related by a dispersive integral in the complex energy domain. As we will now see, for the time-ordered contribution $F_{++}$, the infinite tower of poles cannot be evaded by the integration contour, and will result in series solution for the correlator.

\subsection{Bootstrapping via the spectral representation}
\label{subsec: Bootstrapping via the spectral representation}

We now explicitly compute the integral
\begin{equation}
    F_{++} = - \int_{-\infty}^{+\infty} \d\nu \, \N_\nu \, \frac{F^{(3)}(u^{-1}, \nu) F^{(3)}(v^{-1}, \nu)}{(\nu^2-\mu^2)_{i\epsilon}}\,.
\end{equation}
This spectral representation makes it explicit that the time-ordered contribution $F_{++}$ is the spectral integral of the off-shell function $F_{+-}$ given in Eq.~(\ref{eq: F_+- contribution}). In practice, it means that solutions for the time-ordered contributions are of higher-transcendentality than the factorised ones. We expect that the result contains a local and non-local contributions which take a simple resummed form, as well as an EFT series contribution coming from infinite towers of poles. Proceeding the same way as previously for the three-point function, we decompose the functions $F^{(3)}$ in the Legendre $Q$ basis to isolate well-behaved asymptotic forms at large $\nu$. The integrand takes the form
\begin{equation}
    \begin{aligned}
        \N_\nu \, &F^{(3)}(u^{-1}, \nu) F^{(3)}(v^{-1}, \nu) = \frac{-\nu}{2\pi\sinh(\pi\nu)} \\
        &\times \left[Q_{-i\nu-1/2}(u^{-1})Q_{-i\nu-1/2}(v^{-1}) - Q_{-i\nu-1/2}(u^{-1})Q_{+i\nu-1/2}(v^{-1}) + (\nu \leftrightarrow-\nu)\right]\,.
    \end{aligned}
\end{equation}
First, we immediately observe that both terms related by the shadow transformation $\nu \leftrightarrow -\nu$ contribute equally to the integral. Thus, it suffices to evaluate only one of these terms. Second, we notice that that the term $Q_{-i\nu-1/2}(u^{-1})Q_{-i\nu-1/2}(v^{-1})$ has a well-defined large-$\nu$ behaviour that is independent of the hierarchy between the kinematic variables $u$ and $v$. However, the choice of whether to close the contour upwards of downwards for the term $Q_{-i\nu-1/2}(u^{-1})Q_{+i\nu-1/2}(v^{-1})$ depends on the relative magnitude between $u$ and $v$. As such, we will obtain a solution for $|u|<|v|$ and another one for $|u|>|v|$. The matching condition at $u = v$ is inherently satisfied, as it is encoded in the spectral integral. Therefore, we decompose the spectral integral into two pieces
\begin{equation}
    F_{++} = F_{++}^0 + F_{++}^{<, >}\,,
\end{equation}
with (we have accounted for the factor $2$ coming from the $\nu\leftrightarrow -\nu$ symmetry)
\begin{equation}
    \begin{aligned}
        F_{++}^0 &= \frac{1}{\pi} \int_{-\infty}^{+\infty} \frac{\d\nu\,\nu}{\sinh(\pi\nu)} \frac{Q_{-i\nu-1/2}(u^{-1}) Q_{-i\nu-1/2}(v^{-1})}{(\nu^2-\mu^2)_{i\epsilon}}\,, \\
        F_{++}^{<, >} &= \frac{-1}{\pi} \int_{-\infty}^{+\infty} \frac{\d\nu\,\nu}{\sinh(\pi\nu)} \frac{Q_{-i\nu-1/2}(u^{-1}) Q_{+i\nu-1/2}(v^{-1})}{(\nu^2-\mu^2)_{i\epsilon}}\,.
    \end{aligned}
\end{equation}
Despite the appearance, the contribution $F_{++}^{<, >}$ is well symmetric under $u\leftrightarrow v$, which can be explicitly checked after changing variables $\nu \leftrightarrow -\nu$. The analytic structure of these integrands reveals the presence of several poles with different origins:
\begin{itemize}
    \item A first tower of poles comes from the factor $\nu/\sinh(\pi\nu) = \Gamma(1+i\nu) \Gamma(1-i\nu)/\pi$. They are located on the entire imaginary axis $\nu = \pm i\, (1+n)$ with $n\geq 0$, and cannot be evaded by the contour prescription.
    \item The factor $Q_{-i\nu-1/2}(u^{-1}) Q_{-i\nu-1/2}(v^{-1})$ also generates an infinite tower of (second-order) poles located on the imaginary axis $\nu=-i\, (\tfrac{1}{2}+n)$ with $n\geq0$. However, at large $\nu$, the integrand behaves as $\sim e^{+i\nu(\xi_u+\xi_v)}$ with $\xi_u = \cosh^{-1}(u^{-1}) \geq 0$ for $u^{-1}\geq1$ (and similarly for $\xi_v$), so the contour should be closed in the upper-half complex plane. As a result, we do not collect these poles. This case is similar to the off-shell three-point function case seen in Sec.~\ref{subsec: Off-shell three-point function}. Here, both off-shell exchanged modes interfere constructively so that their spectrum spans only over half the imaginary axis, with residues doubling their weight. 
    \item Poles of the term $Q_{-i\nu-1/2}(u^{-1}) Q_{+i\nu-1/2}(v^{-1})$ are located at $\nu=\pm i\, (\tfrac{1}{2}+n)$ with $n\geq0$. At large $\nu$, it behaves as $\sim e^{+i\nu(\xi_u-\xi_v)}$. For $|u|<|v|$, we have $\xi_u-\xi_v\geq 0$ and we close the contour in the upper-half complex plane, picking the tower of (first-order) poles. This case reflects the destructive interference between off-shell exchanged modes, resulting in the spectrum spanning over the entire imaginary axis. 
    \item Eventually, we also have the usual particle production poles coming from the $i\epsilon$ prescription.
\end{itemize}
Let us now collect these various poles in turn.

\paragraph{Particle production residues.} The poles coming from the $i\epsilon$ prescription are the easiest to evaluate. We obtain
\begin{equation}
    F_{++}^0 \supset \frac{-i}{2\sinh^2(\pi\mu)} \left[e^{+\pi\mu} Q_{+i\mu-1/2}(u^{-1})Q_{+i\mu-1/2}(v^{-1}) + (\mu \leftrightarrow -\mu)\right]\,.
\end{equation}
For the contribution $F_{++}^<$ for which $|u|<|v|$, collecting both poles on the upper-half complex plane results in
\begin{equation}
    F_{++}^< \supset \frac{i}{2\sinh^2(\pi\mu)} \left[e^{+\pi\mu} Q_{+i\mu-1/2}(u^{-1})Q_{-i\mu-1/2}(v^{-1}) + (\mu \leftrightarrow -\mu)\right]\,.
\end{equation}
Combining both contributions, and projecting back onto the Legendre $P$ basis using~(\ref{eq: Legendre P/Q connection formula}) and~(\ref{eq: Legendre P analytic continuation}) yields
\begin{equation}
\label{eq: dS 4pt collider signal}
    F_{++}^{\text{collider}} = - F^{(3)}(-u^{-1}, \mu) F^{(3)}(v^{-1}, \mu)\,,
\end{equation}
for $|u|<|v|$. The contribution for which $|u|>|v|$ is simply found after swapping $u$ and $v$. The found factorised form is not surprising as in the late-time limit time ordering vanishes, as seen in Sec.~\ref{subsec: Spectral representation of massive cosmological propagators}. The result~(\ref{eq: dS 4pt collider signal}) can be directly found by performing the time integrals after substituting $\G_{++} \to \G_{-+}$.

\paragraph{EFT residues.} Contributions coming from EFT residues are analytic in both $u$ and $v$ as $u, v\to0$, and therefore admits series representations. We first sum over the residues from the factor $1/\sinh(\pi\nu)$. The integration contour picks up only half of this tower of first-order poles on the imaginary axis. Explicitly, we obtain
\begin{equation}
\label{eq: F_{++}^0 EFT 1}
    F_{++}^0 \supset -\frac{2}{\pi} \, \sum_{n=0}^{+\infty} (-1)^n \, \frac{(n+1)}{(n+1)^2+\mu^2} \, Q_{n+1/2}(u^{-1}) Q_{n+1/2}(v^{-1})\,,
\end{equation}
where we have used that the residue of $\Gamma(1+i\nu)$ at $\nu = +i(1+n)$ is $\tfrac{(-1)^n}{n!}$. The terms in $1/\mu^2$, characteristic of an EFT expansion, come from 
\begin{equation}
    \frac{1}{(\nu^2-\mu^2)_{i\epsilon}} \xrightarrow[\nu \to +i(n+1)]{} \frac{-1}{(n+1)^2 - \mu^2}\,.
\end{equation}
The large $n$ behaviour of the series coefficients is given by
\begin{equation}
    \left|\frac{(n+1)}{(n+1)^2+\mu^2} \, Q_{n+1/2}(u^{-1}) Q_{n+1/2}(v^{-1})\right| \sim \frac{1}{(n+1)^2+\mu^2} \frac{e^{-(n+1)(\xi_u+\xi_v)}}{\sqrt{\sinh\xi_u \sinh\xi_v}}\,,
\end{equation}
which is well convergent. Notice that close to $u, v=1$, the exponential damping becomes less efficient and the series converges slower, i.e.~as $1/n^2$. Similarly, the contribution of these residues to $F_{++}^<$ is found to be
\begin{equation}
\label{eq: F_{++}^< EFT 1}
    F_{++}^< \supset \frac{2}{\pi} \, \sum_{n=0}^{+\infty} (-1)^n \, \frac{(n+1)}{(n+1)^2+\mu^2} \, Q_{n+1/2}(u^{-1}) Q_{-n-3/2}(v^{-1})\,.
\end{equation}
Quite interestingly, the two towers of residues~(\ref{eq: F_{++}^0 EFT 1}) and~(\ref{eq: F_{++}^< EFT 1}) exactly cancel each other. Indeed, both contributions can be combined in a single series, and using~(\ref{eq: Legendre P/Q connection formula}), each series coefficient is proportional to
\begin{equation}
    Q_{-n-3/2}(v^{-1}) - Q_{n+1/2}(v^{-1}) \propto \sin[\pi(n+1)]\,.
\end{equation}
This effect, which generally occurs in odd spatial dimensions, was previously noted in~\cite{Sleight:2019mgd} and is applicable to contact diagrams as well. Notably, while this phenomenon was initially understood as interference between the two branches of the in-in contour, we show here that this cancellation happens within a single branch. We finish with the poles located at $\nu= \pm i\,(\tfrac{1}{2}+n)$ which only enter the contribution $F_{++}^{<, >}$. They are the only ones contributing to the final expression. For $|u|<|v|$, we obtain the rather simple expression
\begin{equation}
\label{eq: F_{++}^< EFT 2}
    F_{++}^{\text{EFT}} = u\, \sum_{n=0}^{+\infty} \, \frac{(-1)^n}{(n+ \tfrac{1}{2})^2+\mu^2} \, \left(\frac{u}{v}\right)^n \, \pFq{2}{1}{\tfrac{1+n}{2}, 1+\tfrac{n}{2}}{\tfrac{3}{2}+n}{u^2} \, \pFq{2}{1}{\tfrac{1-n}{2}, -\tfrac{n}{2}}{\tfrac{1}{2}-n}{v^2}\,.
\end{equation}
The Legendre $Q$ function being not defined for negative integer order, the above expression is found after evaluating the residue of the $\Gamma$ function entering~(\ref{eq: Legendre P Q definitions}).

\begin{figure}[h!]
    \hspace*{-1.5cm}
    \includegraphics[width=1.2\textwidth]{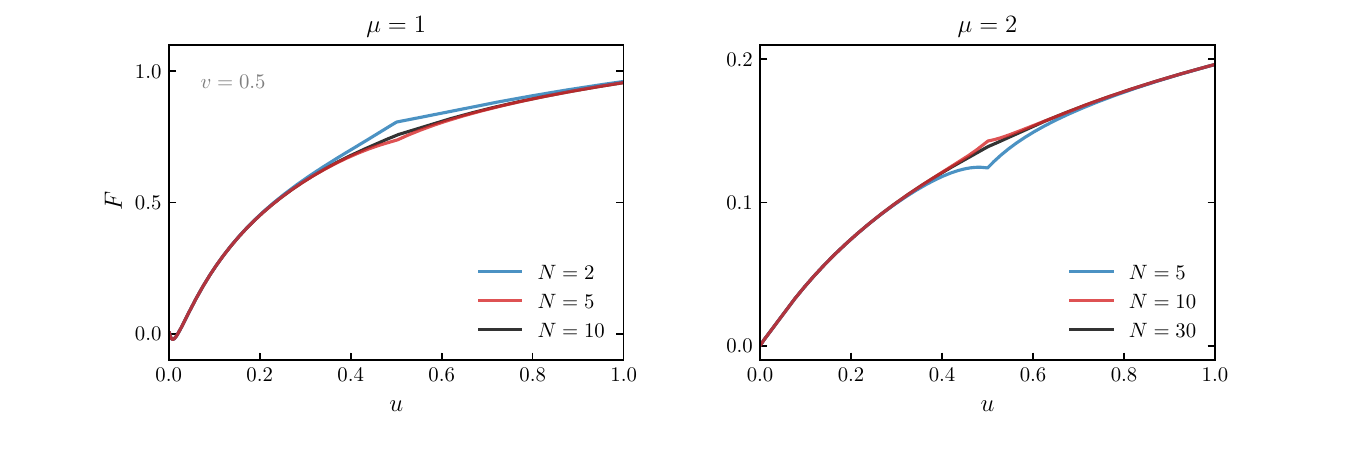}
   \caption{The four-point correlation function $F(u, v) = \sum_{{\sf{a}}, {\sf{b}} = \pm}F_{\sf{ab}}$, as defined in Eq.~(\ref{eq: 4pt function F definition}), as function of the dimensionless kinematic variable $u$ with fixed $v=0.5$, for $\mu=1$ (\textit{left panel}) and $\mu=2$ (\textit{right panel}). We include terms in the series from $n = 1, \ldots, N$. These results reproduce those presented in~\cite{Arkani-Hamed:2018kmz} (Fig.~6) and in~\cite{Qin:2022fbv} (Fig.~6). The \textsf{Mathematica} notebook with the analytical expression for $F$ is available on the Github repository [\href{https://github.com/deniswerth/Spectral-Representation}{\faGithub}].}
  \label{fig: deSitter 4pt function}
\end{figure}

\paragraph{Full result.} The full result is found by summing all contributions, namely the non-time ordered pieces~(\ref{eq: F_+- contribution}) and the time-ordered pieces combining the collider signal~(\ref{eq: dS 4pt collider signal}) with the EFT series~(\ref{eq: F_{++}^< EFT 2}). A few comments are in order. First, this method yields a convergent result in all kinematic configurations. Although not obvious from the analytical expression, we have checked numerically that the found result perfectly matches the one originally found in~\cite{Arkani-Hamed:2018kmz}. Fig.~\ref{fig: deSitter 4pt function} shows the full closed-form result when keeping only a few terms in the series. We observe that the series convergence rate is the same as the bootstrap result but slower than the series solution found using the partial Mellin-Barnes method~\cite{Qin:2022fbv}. Second, compared to the bootstrap result of~\cite{Arkani-Hamed:2018kmz}, the spectral integration naturally resums one of the two series, which in practice makes the evaluation much faster. Lastly, since the spectral representation is solution to the bootstrap equation, there is a one-to-one correspondence between the basis of functions onto which the homogeneous solution of the bootstrap equation is expanded and the Legendre $P$ or $Q$ basis. Additionally, when solving the boundary differential equation, fixing the free coefficients requires two boundary conditions for specific values (or limits) of $u$ and $v$. Here, these coefficients are fully encoded in the spectral integral. Eventually, the spectral method does not requires matching different solutions at $u=v$ because the transition of closing the contour, whether upwards or downwards, is continuous. However, the first derivative of the result is discontinuous at the transition, resulting in a noticeable kink. In short, rather than integrating the boundary differential equation, we directly perform the spectral integral, with the bounds inherently determining the free integration constants.

\paragraph{Largest-time equation.} Assuming $|u|<|v|$, it is trivial to observe that 
\begin{equation}
    F_{++}(u, v) + F_{--}(-u, -v) + F_{+-}(u, -v) + F_{-+}(-u, v) = 0\,,
\end{equation}
from their explicit forms in~(\ref{eq: F_+- contribution}), (\ref{eq: dS 4pt collider signal}) and~(\ref{eq: F_{++}^< EFT 2}). Indeed, both EFT series cancel against each other thanks to the overall factor $u$, i.e.~$F_{--}^{\text{EFT}}(-u, -v) = - F_{++}^{\text{EFT}}(u, v)$. Similarly, $F_{++}^{\text{collider}}(u, v)$ cancels against $F_{-+}(-u, v)$, and $F_{--}^{\text{collider}}(-u, -v)$ against $F_{+-}(u, -v)$. Of course, swapping $u$ and $v$ yields the same result. This provides an additional consistency check for the derived formula.

\subsection{Singularities and analytic structure}

The spectral representation of the four-point function enables straightforward examination of various limits by directly inspecting the spectral integral. A first observation is that removing the dynamical part of the propagating exchanged field---that is, discarding the $i\epsilon$ prescription encoding particle production---, the spectral integral reduces to a contact four-point function generated by the leading bulk interaction $\varphi^4$ in a gradient expansion. This simplification is revealed through the generalised Mehler-Fock transformation (Eq.~(14.20.14) of~\cite{NIST})
\begin{equation}
    \int_{-\infty}^{+\infty} \d\nu \, \N_\nu\, F^{(3)}(u^{-1}, \nu) F^{(3)}(v^{-1}, \nu) = \frac{uv}{u+v}\,.
\end{equation}
This identity is analogous to setting the differential operator of the bootstrap equation to unity and therefore recovering the usual source term. Let us now probe the following limits: (i) collapsed limit $u, v\to0$, i.e.~internal soft momentum $K\to0$ while keeping all external momenta hard, and (ii) the partial-energy pole $u\to -1$ while keeping $v$ fixed.

\paragraph{Collapsed limit.} In the limit $u, v\to 0$, we use the asymptotic behaviour of the Legendre $P$ function~(\ref{eq: Legendre P large z behaviour}) for large argument to easily obtain
\begin{equation}
   \lim_{u, v\to0} F_{+-} = \left(\frac{uv}{4}\right)^{\tfrac{1}{2}+i\mu}\,\frac{\Gamma(\tfrac{1}{2}+i\mu)^2\Gamma(-i\mu)^2}{2\pi}\,.
\end{equation}
Taking the limit inside the spectral integral, we obtain
\begin{equation}
    \lim_{u, v\to0} F_{++} = -\frac{1}{2\pi} \left(\frac{uv}{4}\right)^{\tfrac{1}{2}}\int_{-\infty}^{+\infty} \frac{\d \nu}{(\nu^2-\mu^2)_{i\epsilon}} \left(\frac{uv}{4}\right)^{+i\nu} \Gamma(\tfrac{1}{2}+i\nu)^2\,\frac{\Gamma(-i\nu)}{\Gamma(+i\nu)}\,.
\end{equation}
At large $\nu$, the integrand scales as $\sim e^{i\nu \log(uv/4)}$ (up to a phase) so that the contour is closed in the lower-half complex plane (recall $\log(uv/4)\leq 0$). The usual particle production poles give
\begin{equation}
    \lim_{u, v\to0} F_{++} \supset \frac{i}{2}\left[e^{+\pi\mu}\left(\frac{uv}{4}\right)^{\tfrac{1}{2}+i\mu}\frac{\Gamma(\tfrac{1}{2}+i\mu)^2\Gamma(-i\mu)^2}{2\pi} + (\mu \leftrightarrow-\mu)\right]\,.
\end{equation}
Notice that two infinite towers of poles arise: one from $\Gamma(\tfrac{1}{2}+i\nu)$ located at $\nu = +i\, (\tfrac{1}{2}+n)$ that we do not collect as they lie in the upper-half complex plane, and another from $\Gamma(-i\nu)$ located at $\nu=-i\,n$ with $n\geq 0$ integer, which we do collect. By definition, this EFT signal is analytic in both $u$ and $v$ as it admits a series representation. Its $u, v\to0$ limit can be directly recovered from its full expression~(\ref{eq: F_{++}^< EFT 2}). Eventually, keeping only the terms which are non-analytic in $uv$ as $u, v\to0$ (and therefore also non-analytic in $K$), we obtain
\begin{equation}
    \lim_{u, v\to0} F = \left(\frac{uv}{4}\right)^{\tfrac{1}{2}+i\mu} (1+i\sinh\pi\mu) \, \frac{\Gamma(\tfrac{1}{2}+i\mu)^2\Gamma(-i\mu)^2}{2\pi} + \text{c.c.}\,,
\end{equation}
which exactly reproduces the expression found in~\cite{Arkani-Hamed:2015bza}.

\paragraph{Partial-energy pole.} At fixed $v$, let us probe the limit $u\to -1$. The three-point function $F^{(3)}(z, \mu)$ has a branch point at $z=-1$ so we analyse its behaviour directly at the level of the time integral. The limit $E_1+K\to0$ probes the early time-limit of the left vertex, so we can take the early-time limit of the Hankel function inside the integral
\begin{equation}
    F^{(3)}(u^{-1}, \mu) \sim \frac{1}{\sqrt{2}} \lim_{z\to0}\int_z^{+\infty} \frac{\d x}{x} \,e^{-ix\left(1+u^{-1}\right)} = \frac{1}{\sqrt{2}} \lim_{z\to0} {\rm{E}}_1\left[\left(1+u^{-1}\right)iz\right]\,,
\end{equation}
where ${\rm{E}}_1$ is the exponential integral, which has a logarithmic branch. Isolating the leading term, we obtain
\begin{equation}
    \lim_{u\to-1} F^{(3)}(u^{-1}, \mu) = -\tfrac{1}{\sqrt{2}} \log(1+u)\,.
\end{equation}
The non-time-ordered contributions give
\begin{equation}
     \lim_{u\to-1}(F_{+-} + F_{-+}) = -\sqrt{2}\log(1+u) \times F^{(3)}(v^{-1}, \mu)\,.
\end{equation}
Similarly, the time-ordered contribution is found to be
\begin{equation}
    \lim_{u\to-1} F_{++} = \frac{1}{\sqrt{2}} \log(1+u) \int_{-\infty}^{+\infty}\d\nu\, \N_\nu\, \frac{F^{(3)}(v^{-1}, \nu)}{(\nu^2-\mu^2)_{i\epsilon}} = \frac{1}{\sqrt{2}} \log(1+u) \times F^{(3)}(-v^{-1}, \mu)\,,
\end{equation}
where we have used~(\ref{eq: integrated off-shell 3pt function}). Summing all contributions, we eventually obtain
\begin{equation}
     \lim_{u\to-1}F = -\sqrt{2}\log(1+u) \times \left[F^{(3)}(v^{-1}, \mu) - F^{(3)}(-v^{-1}, \mu)\right]\,.
\end{equation}
In this limit, the residue of the partial-energy pole is proportional to the \textit{shifted} three-point function. This behaviour generalises to any exchanged process where the sum of all ``energies" entering a vertex vanishes, see e.g.~\cite{Arkani-Hamed:2017fdk, Arkani-Hamed:2018bjr, Benincasa:2018ssx}. Similarly, the correlator has a singularity in the flat-space limit $u\to -v$, i.e.~$E = E_1+E_2\to0$ or equivalently $\mu\to\infty$ as seen in Sec.~\ref{subsec: Spectral representation of massive cosmological propagators}, with a coefficient that is related to the flat-space scattering amplitude~\cite{Maldacena:2011nz, Raju:2012zr}.

\newpage
\section{Conclusions}
\label{sec: Conclusions}

In this paper, we have explored an off-shell approach to cosmological correlators. For massive field experiencing particle production, going to the complex mass plane allows one to write a dispersive integral for the propagator that naturally encodes time ordering. Using this object, we have shown that exchanged correlators can be obtained by gluing off-shell lower-point correlators. In this procedure, performing the resulting spectral integral, which amounts to collecting poles as the integrand is meromorphic, effectively sets the exchanged particle on shell. We argued that this representation not only clarifies the analytic structure and factorisation properties of cosmological correlators but also simplifies explicit calculations. As a specific example, we derived a new, simple closed-form formula for the four-point correlator of conformally coupled fields exchanging a massive field in de Sitter space and examined its analytic properties. Additionally, we derived cosmological largest-time equations, which relate individual correlator in-in branch channels through the analytic continuation of certain external energies. These relations, explicitly illustrated for simple diagrams, can serve as consistency checks for more complex correlators.

\vskip 4pt
Eventually, our work opens several interesting avenues for future investigation:

\begin{itemize}
    \item First of all, the spectral representation of massive propagators used in this work does not crucially rely on de Sitter isometries. Meanwhile, a large level of non-Gaussianity can be reached in cosmological processes that strongly break de Sitter boosts. It would therefore be natural to use this spectral representation to obtain clean and simpler closed-form solutions for correlators of particles featuring reduced sound speeds or in the presence of a chemical potential. The latter case is known to boost the particle production rate, which can lead to enhanced cosmological collider signals. However, since mode functions are described by Whittaker functions, the spectral representation should be upgraded accordingly. 
    \item In the context of modern scattering amplitude techniques, complex amplitudes can be efficiently constructed by sewing together lower-point amplitudes using sophisticated recursion relations. We have shown a glimpse of how such procedure can be applied to cosmological correlators in simple cases. Since time integrals are trivialised, computing higher-point correlation functions primarily involves performing a series of spectral integrals, which effectively reduces to summing over several towers of poles, given that off-shell lower-point correlators are meromorphic in the complex mass plane. It will be interesting to explore how this off-shell approach can render computations of more complex correlation functions more tractable. Ultimately, we believe this approach lays the groundwork for deriving more general cosmological recursion relations beyond rational correlators.
    \item Finally, as in flat space, states in a unitary quantum field theory in de Sitter are classified by unitary irreducible representations of the isometries, which are parametrised by the conformal dimension (related to the mass) and spin. Imposing unitarity as an additional consistency condition imposes a set of positivity constraints. While previous such bounds have been largely theoretical, exploring their phenomenological implications could yield valuable insights. The spectral representation employed in this work can be viewed as the leading-order perturbative Källén-Lehmann representation in spatial Fourier space. By perturbatively computing massive self-energy corrections, it may be possible to extend this approach further in perturbation theory, potentially deriving useful bounds from the positivity and meromorphic properties of the spectral density. Ultimately, one would hope to use this spectral representation to ``dress" massive propagators in de Sitter space by resumming bubble diagrams.
\end{itemize}

\paragraph{Acknowledgements.} We thank 
Guillaume Faye,
Jean-Baptiste Fouvry,
Mang Hei Gordon Lee,
Scott Melville,
Enrico Pajer, 
Sébastien Renaux-Petel,
Xi Tong
and 
Zhong-Zhi Xianyu
for helpful discussions. 
We also thank Lucas Pinol, Arthur Poisson and Sébastien Renaux-Petel for comments on the draft.
DW is supported by the European Research Council under the European Union’s Horizon 2020 research and innovation programme (grant agreement No 758792, Starting Grant project GEODESI). 
This article is distributed under the Creative Commons Attribution International Licence (\href{https://creativecommons.org/licenses/by/4.0/}{CC-BY 4.0}).

\newpage
\appendix
\section{Feynman $i\epsilon$ prescription from vacuum wave-functional}
\label{eq: ieps from wavefunctional}

In this appendix, we come back to the $i\epsilon$ prescription defining the in-in generating function~(\ref{eq: SK generating functional}). As is well known, deforming the time integration contour around the infinite past explicitly breaks unitarity~\cite{Feynman:1963fq, Kaya:2018jdo, Baumgart:2020oby, Albayrak:2023hie}. This is a consequence of both in-in branches not being invariant under time reversal, producing a commutator of time evolution operators and not identity. This exact same prescription, when thought of as analytically continuing energy instead of time, not only preserves unitarity, but leads to the correct $i\epsilon$ prescription in the Feynman propagator~(\ref{eq: flat-space iepsilon prescription}).

\vskip 4pt
Reaching convergence in the infinite past is reminiscent of adiabatically projecting the vacuum of the fully interacting theory $\ket{\Omega}$ onto the free one $\ket{0}$. Within the in-in path integral~(\ref{eq: SK generating functional}), this procedure is achieved with the additional term $\Psi[\varphi_+](t_0)\Psi^*[\varphi_-](t_0)$ where $\Psi[\varphi](t) \equiv \braket{\varphi(t)|\Omega}$ is the vacuum wave-functional that describes the transition amplitude from the vacuum $\ket{\Omega}$ to a specific field configuration $\varphi(\bm{x})$ at some time $t_0$~\cite{Weinberg:1996kr, Breuer:2002pc, Calzetta:2008iqa, Chen:2017ryl}. Let us therefore compute this object and take the limit $t_0 \to -\infty$. The key insight is to use the defining property of the annihilation operator $\hat{a}_{\bm{k}}\ket{\Omega} = 0$. After inverting the canonical quantisation of the field $\hat{\varphi}_{\bm{k}}$ and its conjugate momentum $\hat{p}_{\bm{k}}$, and using the Wronskian condition $u_k p_k^* - u_k^*p_k = i$, valid at all time and fixed by the commutation relation ($p_k$ is the mode function associated to $\hat{p}_{\bm{k}}$), one obtains
\begin{equation}
    \hat{a}_{\bm{k}} = \frac{1}{i}\,(p_k^* \hat{\varphi}_{\bm{k}} - u_k^* \hat{p}_{\bm{k}})\,.
\end{equation}
Using the field-space representation of the momentum operator $\hat{p}_{\bm{k}} \to -i\delta/\delta \varphi_{\bm{k}}$, with $\delta \varphi_{\bm{q}}/\delta\varphi_{\bm{k}} = (2\pi)^3 \delta^{(3)}(\bm{q}+\bm{k})$, the condition $\hat{a}_{\bm{k}}\ket{\Omega} = 0$ projected onto $\bra{\varphi(t_0)}$ yields the functional differential equation
\begin{equation}
    \left(u_k^* \frac{\delta}{\delta \varphi_{\bm{k}}} - i p_k^* \varphi_{\bm{k}}\right)\Psi[\varphi](t_0) = 0\,.
\end{equation}
Solving this equation gives the following Gaussian solution for the vacuum wave-functional
\begin{equation}
    \Psi[\varphi](t_0) = \exp\left(-\frac{1}{2}\int\frac{\d^3k}{(2\pi)^3} \varphi_{\bm{k}}(t_0)\omega_k(t_0)\varphi_{-\bm{k}}(t_0)\right)\,,
\end{equation}
where $\omega_k(t_0) = -ip_k^*/u_k^*$, with the mode functions evaluated at $t_0$. The overall normalisation is fixed to unity due to the correct vacuum normalisation $\braket{\Omega|\Omega} = 1$. We now take the limit $t_0\to -\infty$. Reaching the Bunch-Davies state imposes that mode functions behave as simple plane-waves $u_k(t) \sim e^{-ikt}$ so that $\omega_k = k$ is nothing but the dispersion relation. As for the remaining terms, we introduce the following regularisation scheme
\begin{equation}
    f(-\infty) = \underset{\epsilon \rightarrow 0} {\text{lim}} \,\, \epsilon \int_{-\infty}^t \d t' f(t')\,e^{\epsilon t'}\,,
\end{equation}
valid for any smooth-enough function $f(t)$, that can be recovered after integration by parts, to rewrite the contribution from vacuum wave-functionals entering~(\ref{eq: SK generating functional}) as
\begin{equation}
\label{eq: vacuum wave-functional total term}
    \underset{t_0 \rightarrow -\infty} {\text{lim}} \, \Psi[\varphi_+](t_0) \Psi^*[\varphi_-](t_0) = \underset{\epsilon \rightarrow 0} {\text{lim}} \,\, \exp\left(-\frac{\epsilon}{2}\int_{-\infty}^t\d t' \int\frac{\d^3k}{(2\pi)^3}\,k\left[\varphi_{\bm{k}, +}(t') \varphi_{-\bm{k}, +}(t') + (+\leftrightarrow -)\right]\right)\,.
\end{equation}
We have set $e^{\epsilon t'} = 1$ since we only care about the leading term as $\epsilon\to0$. Obtaining a Gaussian weight in the fields with negative real values, as opposed to purely imaginary ones in the standard action, turns out to be crucial to ensure convergence of the path integral in the infinite past. Indeed, let us consider the free quadratic Lagrangian for the fields $\varphi_{\pm}$ written in Fourier space $\mathcal{L}[\varphi_{\bm{k}, \pm}] = -\frac{1}{2}\varphi_{\bm{k}, \pm}(t)(-\partial_t^2 - k^2)\varphi_{-\bm{k}, \pm}(t)$. Adding the term~(\ref{eq: vacuum wave-functional total term}) amounts to modifying the weight in the exponential of the generating function~(\ref{eq: SK generating functional}) to
\begin{equation}
    \pm\frac{i}{2} \int \frac{\d^3k}{(2\pi)^3} \int_{-\infty}^t \d t' \, \varphi_{\bm{k}, \pm}(t')\left[-\partial_{t'}^2 - (k\pm i\epsilon)^2\right]\varphi_{-\bm{k}, \pm}(t')\,,
\end{equation}
where we have used $(k\pm i\epsilon)^2 \approx k^2 \pm 2i\epsilon$ and relabelled $2\epsilon\to \epsilon$. In the end, performing the converging Gaussian path integral over the fields $\varphi_{\pm}$ leads to the correct $i\epsilon$ prescription for the Feynman propagator~(\ref{eq: flat-space iepsilon prescription}).

\section{Details on Massive Fields in de Sitter}
\label{sec: More on massive fields in de Sitter}

Consider a real scalar field $\phi(\tau, \bm{x})$ of mass $m$, where $-\infty<\tau \leq 0$ is the conformal time, evolving in de Sitter space with three spatial
dimensions. We will focus on fields in the principal series so that $\mu \equiv \sqrt{m^2/H^2 - 9/4}$ is real. In what follows, we set $H=1$ for convenience. Defining the canonically normalised field $\sigma(\tau, \bm{x}) \equiv (-\tau)^{-3/2} \phi(\tau, \bm{x})$, its positive-frequency mode function is given by
\begin{equation}
\label{eq: massive dS mode function}
    u_k(\tau, \mu) = \frac{i\sqrt{\pi}}{2} \, e^{-\frac{\pi\mu}{2}} \, H_{i\mu}^{(1)}(-k\tau) \xrightarrow[|k\tau| \to \infty]{} \frac{e^{-ik\tau}}{\sqrt{2k\tau}}\,,
\end{equation}
where $H_{i\mu}^{(1)}(z)$ is the Hankel function of the first kind, such that it reduces to the usual Bunch-Davies vacuum in the far past, up to an overall unimportant phase. The in-in contour $\C_{i\epsilon}$ in~(\ref{eq: SK generating functional}) requires analytically continuing the positive-frequency mode function~(\ref{eq: massive dS mode function}) to the lower-half $k\tau$-complex plane to ensure reaching the asymptotic vacuum. This precisely avoids crossing the branch cut of the Hankel function $H_{i\mu}^{(1)}(z)$, that is chosen to lie on the real negative axis, away from the physical regime. Therefore, using the complex conjugate relation~(\ref{eq: H1-H2 complex conjugate relations}), the negative-frequency mode can be written
\begin{equation}
\label{eq: massive dS mode function neg}
    u_k^*(\tau, \mu) = -\frac{i\sqrt{\pi}}{2} \,e^{+\frac{\pi\mu}{2}}\,H_{i\mu}^{(2)}(-k\tau)\,.
\end{equation}

\subsection{Useful formulae}
\label{subsec: Useful formulae}

The following mathematical identities are useful for manipulating mode functions of a massive scalar field in de Sitter.

\paragraph{Complex conjugate relations.} The Hankel functions of the first and second kind are related by the complex conjugate relations
\begin{equation}
\label{eq: H1-H2 complex conjugate relations}
    \left[H_{i\mu}^{(1)}(z)\right]^* = e^{+\pi\mu}H_{i\mu}^{(2)}(z^*)\,, \quad \left[H_{i\mu}^{(2)}(z)\right]^* = e^{-\pi\mu}H_{i\mu}^{(1)}(z^*)\,,
\end{equation}
for $z \in \mathbb{C}\symbol{92}\{\mathbb{R}^{-}\}$. Notice that these relations are not verified when the argument $z$ lies on the branch cut.

\paragraph{Symmetries.} The Hankel functions obey the following relations
\begin{equation}
\label{eq: symmetry order Hankel}
    H_{-i\mu}^{(1)}(z) = e^{-\pi\mu}H_{i\mu}^{(1)}(z)\,, \quad H_{-i\mu}^{(2)}(z) = e^{+\pi\mu}H_{i\mu}^{(2)}(z)\,,
\end{equation}
for $z \in \mathbb{C}$. These identities render manifest an underlying shadow $\mu \leftrightarrow -\mu$ symmetry for the positive- and negative-frequency mode functions
\begin{equation}
    u_k(\tau, -\mu) = u_k(\tau, \mu)\,, \quad u_k^*(\tau, -\mu) = u_k^*(\tau, \mu)\,.
\end{equation}
Upon rotating time, the Hankel functions satisfy
\begin{equation}
\label{eq: Hankel analytic continuations 1}
    H_{-i\mu}^{(1)}(e^{+i\pi} z) = -H_{i\mu}^{(2)}(z)\,, \quad H_{-i\mu}^{(2)}(e^{-i\pi}z) = -H_{i\mu}^{(1)}(z)\,.
\end{equation}
The first relation is valid for $z$ in the lower-half complex plane excluding the negative real and imaginary axis, and the 
second relation is valid for $z$ in the upper-half complex plane excluding the positive real and imaginary axis. In the physical time domain, these identities relate the mode functions by a CPT symmetry
\begin{equation}
    u_k(z, \mu) = u_k^*(e^{-i\pi}z, -\mu)\,, \quad u_k^*(z, \mu) = u_k(e^{+i\pi}z, -\mu)\,.
\end{equation}
Additional analytic continuations of the Hankel function are given by
\begin{equation}
\label{eq: Hankel analytic continuations 2}
    \begin{aligned}
    H_{i\mu}^{(1)}(e^{-i\pi}z) &= e^{+\pi\mu}H_{i\mu}^{(2)}(z) + 2\cosh(\pi\mu) H_{i\mu}^{(1)}(z)\,, \\ H_{i\mu}^{(2)}(e^{+i\pi}z) &= e^{-\pi\mu}H_{i\mu}^{(1)}(z) + 2\cosh(\pi\mu) H_{i\mu}^{(2)}(z)\,.
    \end{aligned}
\end{equation}

\paragraph{Connection formulae.} Hankel functions can be expanded in terms of Bessel functions with the following connection formulae
\begin{equation}
\label{eq: connection formula}
    H_{i\mu}^{(1)}(z) = \frac{e^{+\pi\mu}J_{+i\mu}(z) - J_{-i\mu}(z)}{\sinh(\pi\mu)}\,, \quad
        H_{i\mu}^{(2)}(z) = \frac{J_{-i\mu}(z) - e^{-\pi\mu}J_{+i\mu}(z)}{\sinh(\pi\mu)}\,.
\end{equation}
The Bessel function of the first kind satisfies 
\begin{equation}
    J_{i\mu}^*(z) = J_{-i\mu}(z^*)\,,
\end{equation}
for $z \in \mathbb{C}\symbol{92}\{\mathbb{R}^{-}\}$, and has the following power-law asymptotic form for small argument $z \to 0$
\begin{equation}
    J_{i\mu}(z) \sim \frac{1}{\Gamma(1+i\mu)} \left(\frac{z}{2}\right)^{i\mu}\,,
\end{equation}
whereas the Hankel function has the following plane-wave asymptotic form for large argument $z\to\infty$
\begin{equation}
    H_{i\mu}^{(1)}(z) \sim \sqrt{\frac{2}{\pi z}}\, e^{i\left(z - \tfrac{\pi}{4}\right) + \tfrac{\pi\mu}{2}}\,.
\end{equation}

\paragraph{Large-order asymptotic forms.} For large $\mu$ with $z$ ($\neq 0$) fixed, we have
\begin{equation}
\label{eq: large-order asymptotic expansion}
    J_{i\mu}(z) \sim \frac{1}{\sqrt{2i\pi\mu}}\left(\frac{ez}{2i\mu}\right)^{i\mu}\,, \quad H_{i\mu}^{(1)}(z) \sim - H_{i\mu}^{(2)}(z) \sim -i \sqrt{\frac{2}{i\pi\mu}} \left(\frac{ez}{2i\mu}\right)^{-i\mu}\,.
\end{equation}

\paragraph{Legendre functions.} The Legendre functions are equivalent to Bessel (and Hankel) functions with higher transcendentality. Explicitly, they are given by
\begin{equation}
\label{eq: Legendre P Q definitions}
    \begin{aligned}
        P_{i\mu-1/2}(z) &\equiv \pFq{2}{1}{\tfrac{1}{2}+i\mu, \tfrac{1}{2}-i\mu}{1}{\frac{1-z}{2}} \,, \\
        Q_{i\mu-1/2}(z) &\equiv \sqrt{\frac{\pi}{2}} \, \frac{\Gamma(\tfrac{1}{2}+i\mu)}{\Gamma(1 + i\mu)} \, 2^{-i\mu} z^{-\tfrac{1}{2}-i\mu} \, \pFq{2}{1}{\frac{3/2+i\mu}{2}, \tfrac{1/2+i\mu}{2}}{1+i\mu}{\frac{1}{z^2}}\,,
    \end{aligned}
\end{equation}
in terms of the hypergeometric function $_2 F_1$ that is defined by the following formal series 
\begin{equation}
    \pFq{2}{1}{a, b}{c}{z} = \sum_{n=0}^\infty \frac{(a)
    _n (b)_n}{(c)_n} \frac{z^n}{n!}\,,
\end{equation}
on the disk $|z|<1$ and by analytic continuation elsewhere, where $(a)_n \equiv \Gamma(a+n)/\Gamma(a)$ is the Pochhammer symbol. These functions are $P_\alpha(z)$ = \textsf{LegendreP[$\alpha$, $0$, $3$, $z$]} and $Q_\alpha(z)$ = \textsf{LegendreQ[$\alpha$, $0$, $3$, $z$]} in Mathematica. The optional argument ``3" selects the correct branch cut structure as it is falsely implemented by default.\footnote{The Legendre functions $P$ and $Q$ generalise to associated Legendre functions (also called Ferrers function of the first and second kind) $P_{i\mu-1/2}(z)\to P_{i\mu-1/2}^{1/2-j}(z)$ and $Q_{i\mu-1/2}(z)\to Q_{i\mu-1/2}^{1/2-j}(z)$, see Chap.~14 of~\cite{NIST}, that are useful when dealing with spatial derivative interactions, i.e.~for different powers of conformal time in the correlator time integrals. The non-derivative interaction case simply corresponds to $j=1/2$.} At fixed argument, the Legendre $P$ function is analytic in the entire complex $\mu$ plane. However, the Legendre $Q$ function is meromorphic as it has poles coming from the factor $\Gamma(\tfrac{1}{2}+i\mu)$.

\paragraph{Integral formula.} The Legendre function $P$ is the solution of the following integrals
\begin{equation}
\label{eq: integral of Hankel}
    \begin{aligned}
        -i \frac{\sqrt{\pi}}{2}e^{+\tfrac{\pi\nu}{2}}\int_{-\infty^+}^0 \frac{\d\tau\, e^{i k\tau}}{(-\tau)^{1/2}} \,  H_{i\nu}^{(2)}(-s\tau) &= \frac{e^{-\tfrac{i\pi}{4}}}{\sqrt{2s}} |\Gamma(\tfrac{1}{2}+i\nu)|^2 \, P_{i\nu-1/2}(k/s)\,, \\
        +i \frac{\sqrt{\pi}}{2}e^{-\tfrac{\pi\nu}{2}}\int_{-\infty^-}^0 \frac{\d\tau \, e^{-i k\tau}}{(-\tau)^{1/2}} \, H_{i\nu}^{(1)}(-s\tau) &= \frac{e^{+\tfrac{i\pi}{4}}}{\sqrt{2s}} |\Gamma(\tfrac{1}{2}+i\nu)|^2 \, P_{i\nu-1/2}(k/s)\,.
    \end{aligned}
\end{equation}

\paragraph{Connection formula.} Analogously to Bessel and Hankel functions, the Legendre functions satisfy the connection formula
\begin{equation}
\label{eq: Legendre P/Q connection formula}
    |\Gamma(\tfrac{1}{2}+i\mu)|^2 \, P_{i\mu-1/2}(z) = \frac{e^{-\tfrac{i\pi}{2}}}{\sinh(\pi\mu)}\left[Q_{-i\mu-1/2}(z) - Q_{+i\mu-1/2}(z)\right]\,.
\end{equation}
Its analytic continuation is defined by
\begin{equation}
\label{eq: Legendre P analytic continuation}
    |\Gamma(\tfrac{1}{2}+i\mu)|^2 \, P_{i\mu-1/2}(z e^{+i\pi}) = \frac{1}{\sinh(\pi\mu)} \left[e^{+\pi\mu}Q_{+i\mu-1/2}(z) - e^{-\pi\mu} Q_{-i\mu-1/2}(z)\right]\,.
\end{equation}

\paragraph{Large-order asymptotic forms.} The Legendre $P$ function has a simple behaviour at $z\to 0$ (similar to Bessel $J$ function), while the Legendre $Q$ function has a simple behaviour at $z\to \infty$ (similar to Hankel $H$ function). Indeed, they are given by
\begin{equation}
    \begin{aligned}
        P_{i\nu-1/2}(\cosh\xi) &\sim \left(\frac{\xi}{\sinh\xi}\right)^{1/2} \,, \\
        Q_{i\nu-1/2}(\cosh\xi) &\sim\frac{\pi}{2i}\left(\frac{2\xi}{\pi\nu\xi\sinh\xi}\right)^{1/2}  e^{-i\left(\nu\xi - \tfrac{\pi}{4}\right)}\,.
    \end{aligned}
\end{equation}
These expressions equivalently give the large-order asymptotic forms. The asymptotic behaviour of the Legendre $P$ function reads
\begin{equation}
\label{eq: Legendre P large z behaviour}
    P_{i\nu-1/2}(z) \sim \frac{1}{\sqrt{\pi}} \frac{\Gamma(-i\nu)}{\Gamma(\tfrac{1}{2}-i\nu)} \left(\frac{1}{2z}\right)^{\tfrac{1}{2}+i\nu}\,,
\end{equation}
for $z\to \infty$.

\subsection{Derivation of the spectral representation}
\label{subsec: Derivation of the spectral representation}

For completeness, we present the proof of the spectral representation of massive scalar field in de Sitter, closely following~\cite{Melville:2024ove}. Let us first consider the case where the time $\tau_1$ is in the future of $\tau_2$, i.e.~$-k\tau_1<-k\tau_2$, so that the time-ordered propagator reduces to $\G_{++}(k; \tau_1, \tau_2) = u_k(\tau_1, \mu) u_k^*(\tau_2, \mu)$. Expanding the first Hankel function $H_{i\nu}^{(2)}(-k\tau_1)$ in~(\ref{eq: massive propagator in de Sitter}) in terms of Bessel functions using~(\ref{eq: connection formula}), using Eq.~(\ref{eq: symmetry order Hankel}), and the symmetry of the pole prescription with respect to $\nu \leftrightarrow -\nu$, the integral can be written
\begin{equation}
    \G_{++}(k; \tau_1, \tau_2) = \frac{i}{2} \int_{-\infty}^{+\infty} \d \nu \,\frac{\nu J_{i\nu}(-k\tau_1) H_{i\nu}^{(2)}(-k\tau_2)}{(\nu^2-\mu^2)_{i\epsilon}}\,.
\end{equation}
The crucial stage is to observe that the numerator $\nu J_{i\nu}(-k\tau_1) H_{i\nu}^{(2)}(-k\tau_2)$ is analytic in the entire complex $\nu$ plane, so that we choose to close the integration contour in the lower-half complex plane~\cite{GUTIERREZTOVAR2007359}. Since the large-$\nu$ expansion~(\ref{eq: large-order asymptotic expansion}) gives
\begin{equation}
    \nu J_{i\nu}(z_1) H_{i\nu}^{(2)}(z_2) \sim \frac{e^{i\nu \log(z_1/z_2)}}{\pi}\,,
\end{equation}
the arc at infinity does not contribute for $z_1<z_2$ and $\text{Im}(\nu) \to -\infty$. By Cauchy's residue theorem, only the two poles located at $\nu = \pm \mu -i\epsilon$ are selected by the closed contour integral, which yields
\begin{equation}
    \begin{aligned}
        \G_{++}(k; \tau_1, \tau_2) &= \frac{\pi}{4\sinh(\pi\mu)} \left(e^{+\pi\mu}J_{+i\mu}(-k\tau_1) H_{+i\mu}^{(2)}(-k\tau_2) - e^{-\pi\mu} J_{-i\mu}(-k\tau_1) H_{-i\mu}^{(2)}(-k\tau_2)\right)\\
        &= u_k(\tau_1, \mu) u_k^*(\tau_2, \mu)\,,
    \end{aligned}
\end{equation}
where we have again used Eqs.~(\ref{eq: symmetry order Hankel}) and~(\ref{eq: connection formula}). For the case where the time $\tau_2$ is in the future of $\tau_1$, i.e.~$-k\tau_2<-k\tau_1$, one needs to expand the second Hankel function $H_{i\nu}^{(2)}(-k\tau_2)$ in~(\ref{eq: massive propagator in de Sitter}) in terms of Bessel functions and then close the contour, picking up the corresponding residues.

\subsection{Non-analyticity and dispersive integral in the energy domain}
\label{subsec: app dispersive integral}

In Sec.~\ref{subsec: Off-shell three-point function}, by performing the spectral integral in the mass domain, we showed that the effect of the particle production poles on the off-shell three-point function is to rotate its external energy $F^{(3)}(z, \mu) \to F^{(3)}(ze^{+i\pi}, \mu)$. Similarly, in Sec.~\ref{subsec: Bootstrapping via the spectral representation} for the four-point function, we have observed that both the factorised contribution $F_{-+}$ and the collider contribution of the nested channel $F_{++}^{\text{collider}}$ are connected through the analytic continuation of external energies. In the following, we show that this connection is manifested through a dispersive integral in the energy domain~\cite{Liu:2024xyi}.

\paragraph{Complex energy domain.} Let us define the off-shell three-point function by its integral representation only
\begin{equation}
    F^{(3)}(z, \mu) \equiv -i \frac{\sqrt{\pi}}{2} \, e^{\tfrac{\pi\mu}{2}+ \tfrac{i\pi}{4}} \sqrt{K} \, \int_{-\infty}^0 \frac{\d\tau\, e^{iE \tau}}{(-\tau)^{1/2}} \,  H_{i\mu}^{(2)}(-K\tau)\,,
\end{equation}
where $z \equiv E/K$, with $E$ and $K$ denoting the external and internal energies, respectively. The physical kinematic region is given by $z \geq 1$. We now want to analytically continue this function in the entire complex $z$ plane, and in particular outside the physical region, without actually evaluating the integral explicitly (the result is given in~(\ref{eq: 3pt function definition})). First, from the asymptotic form of the Hankel function, we notice that the UV convergence as $\tau \to -\infty$ is controlled by the phase of $E+K$, i.e.~the sum of the energies entering this three-point function. The integral converges only when $E+K$ has a negative imaginary part. As well explained in~\cite{Liu:2024xyi}, the integral can be regularised by deforming the contour $-\infty \to -\infty^+(E+K)^{-1}$, with $-\infty^+ \equiv -\infty(1-i\epsilon)$, so that the lower bound $\tau=-\infty$ is approached from a direction that depends on the argument of $E+K$. Then, we note that the integrand exhibits a branch cut, from the Hankel function and the square root, that we choose to be on the positive real axis $\tau \in (0, +\infty)$, as usual. This gives rise to a discontinuity in the energy domain along $E+K \in (-\infty, 0)$, equivalently along $z \in (-\infty, -1)$.

\paragraph{Discontinuity.} We now show that the full function $F^{(3)}(z, \mu)$ can be recovered from the knowledge of its discontinuity only. By slightly deforming the time integration, the discontinuity of the integral along the branch cut $z \in (-\infty, -1)$ is given by the discontinuity of the integrand itself
\begin{equation}
    \begin{aligned}
        \underset{z}{\Disc} \left[F^{(3)}(z, \mu)\right] &= \underset{\epsilon\to0}{\text{lim}}\left[F^{(3)}(ze^{-i\epsilon}, \mu) - F^{(3)}(ze^{+i\epsilon}, \mu)\right] \\
        &= i \frac{\sqrt{\pi}}{2} \, e^{\tfrac{\pi\mu}{2}+ \tfrac{i\pi}{4}} \sqrt{K} \, \int_{\infty}^0\d\tau\, \underset{\tau}{\Disc}\left[ \frac{e^{iE \tau}}{\tau^{1/2}} \,  H_{i\mu}^{(2)}(K\tau)\right]\,.
    \end{aligned}
\end{equation}
Using the known analytic continuation of the Hankel function~(\ref{eq: Hankel analytic continuations 1}) and~(\ref{eq: Hankel analytic continuations 2}), the discontinuity of the integrand is given by
\begin{equation}
    \underset{\tau}{\Disc}\left[ \frac{e^{iE \tau}}{\tau^{1/2}} \,  H_{i\mu}^{(2)}(K\tau)\right] = -2i\cosh(\pi\mu) \, \frac{e^{iE \tau}}{\tau^{1/2}} \,  H_{i\mu}^{(2)}(K\tau)\,.
\end{equation}
Eventually, after changing variables, we obtain
\begin{equation}
\label{eq: discontinuity 3pt}
    \underset{z}{\Disc} \left[F^{(3)}(z, \mu)\right] = -2i\cosh(\pi\mu) \, F^{(3)}(-z, \mu)\,.
\end{equation}

\begin{figure}[h!]
   \hspace*{0.5cm}
    \includegraphics[width=0.5\textwidth]{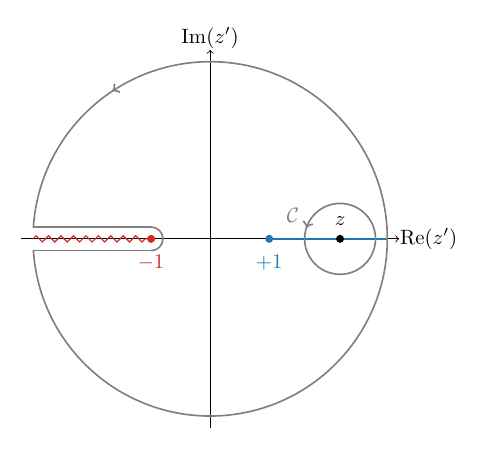}
   \caption{Illustration of the contour deformation used in~(\ref{eq: Cauchy formula for 3pt}) in the complex $z'$ plane. At fixed mass $\mu$, the off-shell three-point function $F^{(3)}(z, \mu)$ in the \textcolor{pyblue}{physical kinematic region $z\geq 1$} is related to its discontinuity along the \textcolor{pyred}{branch cut $z\leq -1$} through a dispersive integral. At a fixed internal energy $K$ flowing in the off-shell leg, the point $z=+1$ is the folded limit $E\to K$ whereas the branch point $z=-1$ is the standard early-time energy singularity $E+K\to0$.}
  \label{fig: dispersive integral complex plane}
\end{figure}

\paragraph{Dispersive integral.} At fixed $\mu$, since the function $F^{(3)}(z, \mu)$ is analytic everywhere except on the branch cut, we can use Cauchy's integral formula to write
\begin{equation}
\label{eq: Cauchy formula for 3pt}
    F^{(3)}(z, \mu) = \oint_{\C} \frac{\d z'}{2i\pi} \frac{F^{(3)}(z', \mu)}{z'-z} = \int_{-\infty}^{-1} \frac{\d z'}{2i\pi} \frac{\Disc_{z'}\left[F^{(3)}(z', \mu)\right]}{z'-z}\,,
\end{equation}
where the integral contour $\C$ encircles $z$ counterclockwise. The second equality is found by deforming the contour as shown in Fig.~\ref{fig: dispersive integral complex plane}, as the integral along the large arc vanishes. Using the found discontinuity of the three-point function~(\ref{eq: discontinuity 3pt}), we finally get
\begin{equation}
    F^{(3)}(z, \mu) = -2i\cosh(\pi\mu) \, \int_{-\infty}^{-1} \frac{\d z'}{2i\pi} \frac{F^{(3)}(-z', \mu)}{z'-z}\,.
\end{equation}
Inversely, $F^{(3)}(-z, \mu)$ is obtained from $F^{(3)}(z, \mu)$ in the same way. The same approach can be employed to reconstruct the complete four-point correlator, including the tower of EFT poles, using only the non-local signal. This powerful method was used and further explored in~\cite{Liu:2024xyi}.

\newpage
\phantomsection
\addcontentsline{toc}{section}{References}
\small
\bibliographystyle{utphys}
\bibliography{references}

\end{document}